\PassOptionsToPackage{unicode}{hyperref}
\PassOptionsToPackage{hyphens}{url}
\PassOptionsToPackage{dvipsnames,svgnames,x11names}{xcolor}
\documentclass[12pt,letterpaper]{article}

\usepackage{amsmath,amssymb}
\usepackage{iftex}
\ifPDFTeX
  \usepackage[T1]{fontenc}
  \usepackage[utf8]{inputenc}
  \usepackage{textcomp} %
\else %
  \usepackage{unicode-math}
  \defaultfontfeatures{Scale=MatchLowercase}
  \defaultfontfeatures[\rmfamily]{Ligatures=TeX,Scale=1}
\fi
\usepackage{lmodern}
\ifPDFTeX\else  
\fi
\IfFileExists{upquote.sty}{\usepackage{upquote}}{}
\IfFileExists{microtype.sty}{%
  \usepackage[]{microtype}
  \UseMicrotypeSet[protrusion]{basicmath} %
}{}
\makeatletter
\@ifundefined{KOMAClassName}{%
  \IfFileExists{parskip.sty}{%
    \usepackage{parskip}
  }{%
    \setlength{\parindent}{0pt}
    \setlength{\parskip}{6pt plus 2pt minus 1pt}}
}{%
  \KOMAoptions{parskip=half}}
\makeatother
\usepackage{xcolor}
\makeatletter
\ifx\paragraph\undefined\else
  \let\oldparagraph\paragraph
  \renewcommand{\paragraph}{
    \@ifstar
      \xxxParagraphStar
      \xxxParagraphNoStar
  }
  \newcommand{\xxxParagraphStar}[1]{\oldparagraph*{#1}\mbox{}}
  \newcommand{\xxxParagraphNoStar}[1]{\oldparagraph{#1}\mbox{}}
\fi
\ifx\subparagraph\undefined\else
  \let\oldsubparagraph\subparagraph
  \renewcommand{\subparagraph}{
    \@ifstar
      \xxxSubParagraphStar
      \xxxSubParagraphNoStar
  }
  \newcommand{\xxxSubParagraphStar}[1]{\oldsubparagraph*{#1}\mbox{}}
  \newcommand{\xxxSubParagraphNoStar}[1]{\oldsubparagraph{#1}\mbox{}}
\fi
\makeatother

\usepackage{longtable,booktabs,array}
\usepackage{calc} %
\usepackage{etoolbox}
\makeatletter
\patchcmd\longtable{\par}{\if@noskipsec\mbox{}\fi\par}{}{}
\makeatother
\IfFileExists{footnotehyper.sty}{\usepackage{footnotehyper}}{\usepackage{footnote}}
\makesavenoteenv{longtable}
\usepackage{graphicx}
\makeatletter
\def\maxwidth{\ifdim\Gin@nat@width>\linewidth\linewidth\else\Gin@nat@width\fi}
\def\maxheight{\ifdim\Gin@nat@height>\textheight\textheight\else\Gin@nat@height\fi}
\makeatother
\setkeys{Gin}{width=\maxwidth,height=\maxheight,keepaspectratio}
\makeatletter
\def\fps@figure{htbp}
\makeatother

\makeatletter
\@ifpackageloaded{caption}{}{\usepackage{caption}}
\AtBeginDocument{%
\ifdefined\contentsname
  \renewcommand*\contentsname{Table of contents}
\else
  \newcommand\contentsname{Table of contents}
\fi
\ifdefined\listfigurename
  \renewcommand*\listfigurename{List of Figures}
\else
  \newcommand\listfigurename{List of Figures}
\fi
\ifdefined\listtablename
  \renewcommand*\listtablename{List of Tables}
\else
  \newcommand\listtablename{List of Tables}
\fi
\ifdefined\figurename
  \renewcommand*\figurename{Figure}
\else
  \newcommand\figurename{Figure}
\fi
\ifdefined\tablename
  \renewcommand*\tablename{Table}
\else
  \newcommand\tablename{Table}
\fi
}
\@ifpackageloaded{float}{}{\usepackage{float}}
\floatstyle{ruled}
\@ifundefined{c@chapter}{\newfloat{codelisting}{h}{lop}}{\newfloat{codelisting}{h}{lop}[chapter]}
\floatname{codelisting}{Listing}

\makeatother
\makeatletter
\@ifpackageloaded{caption}{}{\usepackage{caption}}
\@ifpackageloaded{subcaption}{}{\usepackage{subcaption}}
\makeatother

\ifLuaTeX
  \usepackage{selnolig}  %
\fi
\usepackage[]{natbib}
\usepackage{bookmark}

\IfFileExists{xurl.sty}{\usepackage{xurl}}{} %
\usepackage{placeins}
\usepackage{amsthm,mathtools}
\usepackage{tabularx,enumitem}
\usepackage{tikz}
\usepackage[capitalise,noabbrev]{cleveref}
\usepackage[acronym,nonumberlist]{glossaries}

\newacronym{AIS}{AIS}{adaptive importance sampling}

\newacronym{CRN}{CRN}{common random number}
\newacronym{CDF}{CDF}{cumulative distribution function}

\newacronym{ESS}{ESS}{effective sample size}

\newacronym{IS}{IS}{importance sampling}

\newacronym{KR}{KR}{Knoethe-Rosenblatt}

\newacronym{MC}{MC}{Monte Carlo}
\newacronym{MSE}{MSE}{mean squared error}
\newacronym{MCMC}{MCMC}{Markov chain Monte Carlo}
\newacronym{MVN}{MVN}{multivariate normal}
\newacronym{MVN-MIX}{MVN-MIX}{multivariate normal mixture}

\newacronym{PDF}{PDF}{probability density function}

\newacronym{SNIS}{SNIS}{self-normalized importance sampling}

\newacronym{UIS}{UIS}{unnormalized importance sampling}

\newacronym{VQ}{VQ}{vector quantile}

\newacronym{TABI}{TABI}{target-aware Bayesian inference}

\newacronym{SVD}{SVD}{singular value decomposition}

\newacronym{VI}{VI}{variational inference}

\newacronym{SMC}{SMC}{sequential Monte Carlo}
\newacronym{BS}{BS}{bridge sampling}
\newacronym{i.i.d.}{i.i.d.}{independent and identically distributed}
\newacronym{AMIS}{AMIS}{adaptive multiple importance sampling}
\glsdisablehyper

\newcommand{\x}{\boldsymbol{x}}

\newcommand{\z}{\boldsymbol{z}}
\newcommand{\n}{\boldsymbol{n}}

\newcommand{\bmu}{\boldsymbol{\mu}}

\newcommand{\thetab}{\boldsymbol{\theta}}

\newcommand{\thetabplain}{\boldsymbol{\theta}}
\newcommand{\muplain}{\boldsymbol{\mu}}
\newcommand{\Sigmaplain}{\boldsymbol{\Sigma}}
\newcommand{\dimens}{d_x}
\newcommand{\SNISest}{\widehat{I}_{\mathrm{SNIS}}^{N}}
\newcommand{\GenSNISest}{\widehat{I}_{\mathrm{GenSNIS}}^{N}}

\newcommand{\hatj}{\widehat{J}^{N}}
\newcommand{\hatz}{\widehat{Z}_{\pi}^{N}}
\newcommand{\jointprop}{\mathbb{Q}_{1:2}}
\newcommand{\proposalspaceone}{\mathcal{Q}_1}
\newcommand{\proposalspacetwo}{\mathcal{Q}_2}
\newcommand{\proposalspaceoneparametrizedplain}{\mathcal{Q}_{\boldsymbol{\theta}_1}}
\newcommand{\proposalspacetwoparametrizedplain}{\mathcal{Q}_{\boldsymbol{\theta}_2}}
\newcommand{\couplingmeasure}{\mathbb{C}_{1:2}(d\boldsymbol{u}_1,d\boldsymbol{u}_2)}
\newcommand{\eqdef}{\stackrel{\mathrm{def}}{=}}
\newcommand{\iidsim}{\stackrel{\mathrm{i.i.d.}}{\sim}}

\providecommand{\abs}[1]{\lvert#1\rvert}
\providecommand{\widebar}[1]{\overline{#1}}

\DeclareMathOperator*{\argmax}{arg\,max}
\DeclareMathOperator*{\argmin}{arg\,min}

\newcommand{\cblue}[1]{#1}

\newcommand{\circled}[1]{\tikz[baseline=(char.base)]{\node[shape=circle,draw,inner sep=2pt] (char) {#1};}}

\newcolumntype{Y}{>{\centering\arraybackslash}X}
\usepackage{aliascnt}
\newtheorem{theorem}{Theorem}[section]
\newaliascnt{proposition}{theorem}
\newtheorem{proposition}[proposition]{Proposition}
\aliascntresetthe{proposition}
\newaliascnt{lemma}{theorem}

\aliascntresetthe{lemma}
\theoremstyle{definition}
\newaliascnt{definition}{theorem}
\newtheorem{definition}[definition]{Definition}
\aliascntresetthe{definition}
\newaliascnt{remark}{theorem}
\newtheorem{remark}[remark]{Remark}
\aliascntresetthe{remark}
\newaliascnt{example}{theorem}
\newtheorem{example}[example]{Example}
\aliascntresetthe{example}
\numberwithin{equation}{section}
\Crefname{equation}{equation}{equations}
\Crefname{equation}{Equation}{Equations}
\Crefname{section}{section}{sections}
\Crefname{section}{Section}{Sections}
\Crefname{subsection}{section}{sections}
\Crefname{subsection}{Section}{Sections}
\Crefname{proposition}{proposition}{propositions}
\Crefname{proposition}{Proposition}{Propositions}
\Crefname{definition}{definition}{definitions}
\Crefname{definition}{Definition}{Definitions}
\Crefname{example}{example}{examples}
\Crefname{example}{Example}{Examples}
\makeatletter
\let\paragraph\oldparagraph
\let\subparagraph\oldsubparagraph
\makeatother
\setcitestyle{authoryear,round,semicolon,aysep={},yysep={,}}
\hypersetup{colorlinks=true,allcolors=red}
\allowdisplaybreaks[1]


\newcommand{\PaperTitle}{Generalizing Self-Normalized Importance Sampling with Couplings}
\newcommand{\PaperAuthors}{Nicola Branchini; V\'{i}ctor Elvira}
\newcommand{\NicolaAffiliation}{Department of Statistics, University of Warwick, UK}
\newcommand{\VictorAffiliation}{School of Mathematics, University of Edinburgh, UK}

\newcommand{\PaperKeywords}{Importance sampling; Bayesian computation; ratio estimation; variance reduction}

\newcommand{\AdditionalAcknowledgments}{}

\newcommand{\ArxivFundingStatement}{}
\newcommand{\ArxivDisclosureStatement}{}
\newcommand{\ArxivDataAvailabilityStatement}{}

\hypersetup{
  pdftitle={\PaperTitle},
  pdfauthor={\PaperAuthors},
  pdfkeywords={\PaperKeywords},
  pdfcreator={LaTeX}}
\def\spacingset#1{\renewcommand{\baselinestretch}{#1}\small\normalsize}
\title{\bfseries \PaperTitle}
\author{Nicola Branchini\textsuperscript{1}\hspace{1.5em}V\'{i}ctor Elvira\textsuperscript{2}\\[0.3em]
  {\small \textsuperscript{1}\NicolaAffiliation}\\
  {\small \textsuperscript{2}\VictorAffiliation}}
\date{}
\begin{document}
\spacingset{1}
\maketitle
\begin{abstract}

Estimating expectations under distributions known up to a normalizing constant is a fundamental challenge in statistics. The self-normalized importance sampling (SNIS) estimator is widely used for its simplicity. However, SNIS suffers from high variance in some applications, such as Bayesian sensitivity or outlier analysis. Further, there is a lack of general methods to select a proposal tailored to SNIS. 
We introduce a generalization of SNIS by exploiting its interpretation as a ratio of two IS estimators. 
We sample from a joint distribution, whose marginals play the role of proposals for numerator and denominator, thereby including existing two-proposal methods as special cases. Our perspective allows us to develop new theory for SNIS. We show that an appropriate selection of the joint can bring the asymptotic variance from infinite to finite.
  We prove sufficient and necessary sample size results for generalized SNIS valid under infinite variance, inspired by Chatterjee \& Diaconis (2018). Differently from existing results: ours further apply for non-square integrable functions, show the role of the dependence between numerator and denominator, and connect with the known optimal SNIS proposal of Hesterberg (1988). We demonstrate the benefits of the proposed methodology across applications in Bayesian sensitivity and outlier analysis.

\end{abstract}
\medskip
\glsresetall
\begingroup
\section{Introduction}\label{sec:intro} 
   \Gls{IS}  is a methodological framework that generalizes the \gls{MC} principle to approximate integrals \citep{agapiou2017importance,elvira2021advances}. These integrals often take the form of an expectation of some test function under a target density.
   In \gls{IS}, integrals are approximated by weighted sums using samples generated by a so-called proposal density.
   Within statistics, \gls{IS} has found numerous applications to Bayesian computation where it has demonstrated state-of-the-art performance combined with \gls{MCMC} and/or neural networks \citep{martino2017layered,DBLP:conf/uai/HanL17,muller2019neural,thin2021neo} for adapting the proposal or in combination with annealing \citep{neal2001annealed}, as well as its extension to sequential MC samplers (SMC) \citep{del2006sequential}.

 In many applications, notably whenever Bayesian posterior predictive values are involved such as in Bayesian cross validation \citep{vehtari2017practical} or Bayesian outlier analysis \citep{peng1995bayesian}, conformal Bayesian inference \citep{fong2021conformal}, probabilistic programming \citep{reicheltrethinking} and a-posteriori rare event estimation \citep{picard2013rare} (among others), \gls{MC} estimators - without appropriate \gls{IS} - can result in unacceptably large variance.
 The de facto standard approach in these problems is to use a particular \gls{IS} estimator, which normalizes the \gls{IS} weights: the \gls{SNIS} estimator.\footnote{Other names for the \gls{SNIS} estimator include autonormalized estimator \citep{agapiou2017importance}, or Hajek estimator \citep{khan2021adaptive}.} 
 Currently, the most widely used \gls{AIS} schemes adapt the proposals without considering that the \gls{SNIS} estimator will be used for the final estimation, e.g., for non-parametric approaches like population Monte Carlo algorithms \citep{cappe2004population,elvira2017improving} or the safe AIS \citep{10.1214/20-AOS1983,korba2022adaptive}, moment-matching schemes like adaptive multiple \gls{IS} (AMIS) \citep{cornuet2012adaptive} or \citep{portier2018asymptotic}, as well as other gradient-based methods \citep{ryu2014adaptive,akyildiz2021convergence,elvira2022optimized,boom2024doubly} (see \citep{bugallo2017adaptive} or \citep{portier2018asymptotic,mcbook} for an overview of existing algorithms). More precisely, all of these algorithms do not take into account the normalization of the weights and design proposals that are close to the target density. This is a fundamental limitation when \gls{SNIS} is used because only an unnormalized target density is available.

 In this work, we will develop a new framework that significantly generalizes the \gls{SNIS} estimator, fully exploiting its view as a ratio estimator: it is the ratio of two unbiased IS estimators, each estimating a different integral; the integral in the numerator contains the test function (see \Cref{subsec:snisback}).
The new framework contains as particular cases many previous works that use two different proposals to estimate the two integrals. We now give an overview of related works, which will also be discussed throughout the paper. 

\textbf{Connections with existing work.} The core idea of using a different proposal for the numerator and denominator integrals appears as early as \cite{goyal1987measure}. A ratio of two independent importance sampling estimators, each with its own proposal, also appears in \citet[Section~2.1]{chen1997monte} in their ratio IS methods for estimating ratios of normalizing constants. More recently, \cite{rainforth2020target} (extending \citep{golinski2018}) explained the limitations of the SNIS estimator for certain Bayesian inference problems, namely those where the underlying quantity of interest is a Bayesian posterior predictive value, and revisited the two proposal idea. They are the first to demonstrate on realistic Bayesian inference problems that substantial variance reduction can be obtained thanks to the flexibility of allowing different proposals for numerator and denominator. However, a key drawback of two-proposal approaches is that they construct independent estimators for the numerator and denominator, while the SNIS estimator can reduce variance by exploiting their correlation. Some previous works have explored the idea of maintaining some form of depedence while allowing different proposals for the numerator and denominator, but they are either limited to specific parametric families of marginals or to specific types of dependence \citep{lamberti2018double,cui2024deep}.

Expectations, in the special case of non-negative test functions, can be seen as a ratio of normalizing constants. There exist specialised methods for these ratios, such as bridge sampling (BS) \citep{meng1996simulating}. BS uses samples from the two distributions and works better when these overlap more, in some sense \citep{mcbook}. Warp bridge sampling \citep{meng2002warp,wang2022warp} attempts to improve this overlap by transforming the distributions while preserving their normalizing constants. \cite{guedon2025estimation} proposes a related recent method for ratios of constants. These ideas are related to the framework we develop, but we deal with the more general problem of expectation estimation, including signed test functions. It is also worth mentioning that the previously mentioned \citep{chen1997monte} already showed that the asymptotic variance of ratio IS methods - closer to our work - is lower than that of optimal bridge sampling, in an i.i.d. setting.

Let us also mention related methods from the wider Monte Carlo literature. \citet{vardi1985empirical} studied a statistical model where several sampling distributions are obtained from the same unknown distribution by applying known nonnegative weighting functions. The aim there is to estimate this common distribution using all the samples. \citet[Sections~3.1--3.2]{kong2003theory} develop this idea for Monte Carlo integration: they estimate the underlying measure by maximum likelihood and then integrate the test function of interest against that estimate. \citet{tan2004likelihood} develops the likelihood approach further.

Another set of relevant approaches consists in retaining the SNIS estimator and choose a flexible proposal that can be sampled only with MCMC \citep{silva2022robust,paananen2021implicitly}, although this complicates theory; these works's analyses do not account for the added autocorrelation. More recently, \citet{owen2025zero} develops an estimating-equation extension of \gls{SNIS} that uses two proposals and whose asymptotic variance can approach zero. We will expand on additional detailed connections to previous works throughout the paper and in \Cref{sec:connecs}.

\textbf{Contributions.} In contrast to these previous works, we will provide a general framework that allows for arbitrary parametric marginals and investigates many flexible forms of dependence between the numerator and denominator estimators. We are also the first to provide a systematic conceptual study of the effect of the dependence between numerator and denominator on the variance of the \gls{SNIS} estimator, in particular its connection to the choice of proposal distribution - along the way, also providing new theoretical insights about how an appropriate choice of dependence can reduce variance from infinite to finite and a certain generalization of the celebrated Chatterjee-Diaconis non-asymptotic results for importance sampling \citep{chatterjee2018sample} Our contributions are detailed below.

\begin{itemize}
   \item We show that the asymptotic variance of the \gls{SNIS} estimator admits an interpretable decomposition with three terms, clearly highlighting the dependence on the proposal density in all three. 
   The third term, which has received significantly less attention w.r.t. the selection of the proposal in \gls{IS}, can be exploited for variance reduction according to the dependence between the numerator and denominator estimators.
   \item We exploit the decomposition above to develop a generalization of the \gls{SNIS} estimator that uses a joint distribution in an extended space to parameterize the dependence between the numerator and denominator estimators. The formulation puts under a unifying framework and considerably generalizes works using two proposals for ratio estimation using \gls{IS}, which either (i) did not introduce dependence, (ii) were restricted to certain parametric families of marginals, or (iii) used restricted types of dependence.  
   \item We develop a construction of the joint distribution by decomposing it into two multivariate marginals, which play the role of interpreted proposals, and a coupling, defined in this paper as a certain generalization of the notion of copula \citep{fan2023vector}. We therefore open up a new framework to design proposal distributions in IS, where adaptation can be carried out both for marginals and for the coupling. 
   \item The perspective of the framework leads to new theory for generalized SNIS. First, we show that a suitable choice of coupling can bring the asymptotic variance from infinite to finite. Then we prove non-asymptotic bounds on the $L_1$ error which lead to sufficient and necessary sample size considerations, inspired by \citep{chatterjee2018sample}. Unlike their results, ours are specific to SNIS as (i) the test function is taken into account (ii) the numerator/denominator dependence plays an explicit role and (iii) the optimal SNIS proposal of \citep{hesterberg1988advances} appears naturally.
   \item We demonstrate the usefulness of our framework in motivated statistical applications - Bayesian outlier analysis and sensitivity analysis - where a good choice of proposal distribution for SNIS is essential. 
\end{itemize} 
 
\paragraph{Paper structure.}
In \Cref{sec:ch2-limitations} we show a concrete setting where \gls{SNIS} is typically used and its limitations. In \Cref{sec:main_methodology} we introduce our generalized \gls{SNIS} framework: \Cref{sec:jointprop_intro} defines the extended-space joint proposal, \Cref{sec:construction_joint} shows how to construct it via couplings and transport maps, \Cref{sec:optimalcoupling} and \Cref{sec:paramcoupling} address optimization and concrete coupling choices, \Cref{sec:theory} gives theoretical guarantees, and \Cref{sec:connecs} connects to prior work. In \Cref{sec:experiments} we demonstrate how to obtain more accurate estimates of expectations in two applications exploiting our generalized SNIS perspective. We conclude in \Cref{sec:exp_discussion} with a discussion and ideas for future work. The Appendix includes proofs and additional details about the experiments. 

  \textbf{Notation and standing assumptions.}
Let $\mathcal X\subseteq\mathbb R^{\dimens}$ be a Borel set. We write $\mathbb P(\mathcal X)$ for its probability measures and $\mathcal P(\mathcal X)$ for the Lebesgue densities of such measures. A measure such as $\mathbb Q$ is written in blackboard bold, its density $q$ in lowercase when one exists, and its distribution function $Q$ in uppercase. For $r\geq1$, $L^r(q)$ contains functions $f$ for which $\int_{\mathcal X}|f(\boldsymbol{x})|^r q(\boldsymbol{x})\,d\boldsymbol{x}<\infty$. For densities $p,q$, the notation $p\ll q$ means that the measure with density $p$ is absolutely continuous w.r.t. the measure with density $q$. If $\boldsymbol{T}$ is measurable, $\boldsymbol{T}\sharp\mathbb Q$ is the distribution of $\boldsymbol{T}(\boldsymbol{x})$ for $\boldsymbol{x}\sim\mathbb Q$.

\subsection{Self-normalized IS and its limitations}\label{subsec:snisback} 
Let $\mathbb P_\pi$ have density $\pi=\widetilde\pi/Z_\pi$, where $\widetilde\pi\geq0$ and $0<Z_\pi=\int_{\mathcal X}\widetilde\pi(\boldsymbol{x})\,d\boldsymbol{x}<\infty$. For a measurable, possibly unbounded test function $\varphi:\mathcal X\to\mathbb R$ with $\varphi\in L^1(\pi)$, our quantity of interest is the expectation
\begin{align}\label{eq_snis_problem}
    I  & \eqdef \mathbb{E}_{\mathbb{P}_{\pi}}[\varphi(\x)] = \int_{\mathcal{X}} \varphi(\x) \mathbb{P}_{\pi}(d \x) = \int_{\mathcal{X}} \varphi(\x) \pi(\x) d\x .
\end{align} 
Typically, $Z_\pi$ is unknown so only $\widetilde \pi$ can be evaluated, not $\pi$. Therefore, ordinary \gls{UIS}, which requires evaluations of $\pi/q$, cannot be used.

\textbf{Self-normalized IS (SNIS).} Let $q$ be a proposal satisfying $\pi\ll q$, and sample $\x^{(1)},\ldots,\x^{(N)}$ i.i.d. from $q$. When the weights have a positive sum, normalize them as $\widebar w^{(n)}=w(\x^{(n)})/\sum_{i=1}^Nw(\x^{(i)})$. The resulting estimator is the \gls{SNIS} estimator, 
\begin{align} \label{eq_snis_est_again}
    \SNISest \eqdef \sum_{n=1}^{N}  \widebar{w}^{(n)} \varphi(\x^{(n)}) , \qquad   \{ \x^{(n)} \}_{n=1}^{N}  \iidsim q(\x) . 
\end{align}
If $q$ samples outside the support of $\widetilde\pi$, all $N$ weights may be zero, an event with probability $q\{\widetilde\pi=0\}^N$: this is fast enough to not affect asymptotic variance (defined later).

 While \Cref{eq_snis_est_again} shows how to compute the estimator, it hides that, implicitly, two integrals are being estimated. Indeed, we can write $I$ as:
\begin{align}\label{eq_snis_ratio}
             I = \int_{\mathcal{X}} \varphi(\x) \pi(\x) d\x =  \frac{\int_{\mathcal{X}} \varphi(\x) \widetilde{\pi}(\x) d\x}{\int_{\mathcal{X}} \widetilde{\pi}(\x) d\x} = \frac{J}{Z_{\pi}},
\end{align}
where we denoted the integral in the numerator as $J$. The observation leads to a more informative expression for $\SNISest$. By defining two \gls{UIS} estimators constructed with a shared set of samples $\{ \x^{(n)} \}_{n=1}^{N} \iidsim q(\x)$, i.e., $ \hatj \eqdef  \frac{1}{N}\sum_{n=1}^{N} \frac{\varphi(\x^{(n)}) \widetilde{\pi}(\x^{(n)})}{q(\x^{(n)})}$ and $\hatz \eqdef \frac{1}{N} \sum_{n=1}^{N} \frac{\widetilde{\pi}(\x^{(n)})}{  q(\x^{(n)})}$,

on the event $\hatz>0$, the \gls{SNIS} estimator can then be written as their ratio:
\begin{align}\label{eq_snis_est}
    \SNISest &= \frac{\hatj}{\hatz} = \frac{\frac{1}{Z_{\pi}} \frac{1}{N}\sum_{n=1}^{N} \frac{\varphi(\x^{(n)}) \widetilde{\pi}(\x^{(n)})}{q(\x^{(n)})}}{\frac{1}{Z_{\pi}} \frac{1}{N} \sum_{n=1}^{N} \frac{\widetilde{\pi}(\x^{(n)})}{  q(\x^{(n)})}} =  \frac{\sum_{n=1}^{N}\frac{\varphi(\x^{(n)})\widetilde{\pi}(\x^{(n)})}{q(\x^{(n)})}}{\sum_{n=1}^{N}\frac{\widetilde{\pi}(\x^{(n)})}{q(\x^{(n)})}}. 
\end{align}
Notice that $\hatz$ is the denominator of $  \widebar{w}^{(n)}$, except for the factor $1/N$ which cancels in the normalization. While \Cref{eq_snis_est}  is mathematically equivalent to \Cref{eq_snis_est_again}, it shows explicitly that SNIS is a ratio of the two \gls{UIS} estimators, $\hatj$ and $\hatz$, which share the same set of samples $\{ \x^{(n)} \}_{n=1}^{N}$. 

\textbf{Selection of the proposal.} As in any \gls{IS} method, the key challenge for SNIS lies in selecting a good proposal $q$. To compare proposals, we will focus on asymptotic variance. \citet{deligiannidis2024importance} instead studies the bias more in detail. We use $\mathbb V_q^\infty[\SNISest]$ for asymptotic variance and we specifically mean the variance of the limiting Gaussian distribution given by the central limit theorem (CLT), $\sqrt N(\SNISest-I)\xrightarrow{d}\mathcal N(0,\mathbb V_q^\infty[\SNISest])$, where\footnote{It is also possible to define asymptotic variance as $\lim_{N\to\infty}N\mathbb V_q[\SNISest]$, when this exists. In general, they are not equal without further moment assumptions.}
$
\mathbb V_q^\infty[\SNISest]=\int_{\mathcal X}\frac{\pi(\boldsymbol{x})^2}{q(\boldsymbol{x})}\{\varphi(\boldsymbol{x})-I\}^2\,d\boldsymbol{x} .
$
Though not often discussed explicitly, as we see in \cref{sec:theory}, we only need $\pi(\varphi-I)/q\in L^2(q)$ for the CLT to hold, not individual moments of the IS weights. In search for a good $q$, we can minimize the asymptotic variance, giving the lower bound \citep{mcbook}
\begin{align}
\label{eq:lb}
\mathbb V_q^\infty[\SNISest]
\geq \bigl(\mathbb{E}_\pi|\varphi-I|\bigr)^2 = || \varphi - I ||_{L_1(\pi)}^2.
\end{align}
While simple, the observation that the minimal variance is the $L_1(\pi)$ norm of $\varphi - I$ is overlooked in the literature. Our applications in \Cref{sec:experiments} will demonstrate that we may well want to integrate $L_1(\pi) \setminus$ functions in practice, and this requires careful choice of $q$. The corresponding optimal proposal is $q_{\mathrm{SNIS}}^\star(\boldsymbol{x})=\pi(\boldsymbol{x})|\varphi(\boldsymbol{x})-I|/\mathbb{E}_\pi|\varphi-I|$ \citep{hesterberg1988advances}.\footnote{The density attains the bound when $\pi\{\varphi=I\}=0$, because otherwise it violates $\pi\ll q$.} This choice depends on the unknown $I$, and even if it were available, the lower bound would remain positive. \citet{rainforth2020target} demonstrate the importance of this limitation: their two-proposal estimators achieve substantially smaller errors than even $q_{\mathrm{SNIS}}^\star(\boldsymbol{x})$ in several examples from Bayesian statistics.

A different approach is to use a flexible proposal known only up to a normalizing constant and obtain samples with \gls{MCMC}. In this spirit, \citet{silva2022robust} construct a mixture proposal for estimating \emph{all} leave-one-out predictive density values from a single chain jointly, whereas \citet{paananen2021implicitly} post-process MCMC samples from the full posterior distribution to bridge them closer to an approximation of $q_{\mathrm{SNIS}}^\star(\boldsymbol{x})$. This is a different goal than ours. Closer to the latter work, we focus on estimating accurately \emph{individual} expectations, i.e., for a specific $\varphi$. The aforementioned flexibility of choosing a flexible proposal to be sampled with MCMC implies that these methods in practice cannot have lower asymptotic variance than i.i.d. SNIS with the same $q$ (since besides toy examples, MCMC autocovariances are non-negative) and thus of the lower bound in \Cref{eq:lb}. 

\subsubsection*{A motivating class of examples \& limitations of SNIS}\label{sec:limits}\label{sec:ch2-limitations}
\cblue{ We present a concrete setting where SNIS is typically used, while its limitations are often overlooked and not laid out explicitly. A similar discussion can be found in \cite{rainforth2020target} to motivate two-proposal, which our framework generalizes. }

\begin{example}[Bayesian cross-validation, outlier analysis]\label{mot:example}

    Test data often comes from a different distribution due to outliers \citep{paananen2021implicitly} or model misspecification: $y_{S+1} \sim g(y; \thetab_{\text{wrong}})$, $\thetab_{\text{wrong}} \neq \thetab_{\bigstar}$,\footnote{The following discussion can be easily generalized for multiple test observations $y_{S+1}, y_{S+2}, \dots$, and multivariate observations.} or $y_{S+1} \sim h(y; \boldsymbol{\gamma})$ for a different model $\{h(y; \boldsymbol{\gamma}) \}_{\boldsymbol{\gamma} \in \Gamma}$ where there is no $\boldsymbol{\gamma} : h(y; \boldsymbol{\gamma}) = g(y; \thetab_{\bigstar})$. For fixed $y_{S+1}$, out-of-sample performance is evaluated by estimating
    \begin{align}\label{eq:post_predictive}
        I = \mathbb{E}_{\pi(\thetab | \boldsymbol{y}_{1:S})}[ g(y_{S+1} | \thetab ) ] = \int_{\Theta} g(y_{S+1} | \thetab ) \pi(\thetab | \boldsymbol{y}_{1:S}) d \thetab = \frac{ \int_{\Theta} g(y_{S+1} | \thetab ) \widetilde{\pi}(\thetab | \boldsymbol{y}_{1:S}) d \thetab}{ \int_{\Theta} \widetilde{\pi}(\thetab | \boldsymbol{y}_{1:S}) d \thetab}. 
    \end{align}
 Such a quantity is of interest in many modern statistical problems, including Bayesian leave-one-out cross-validation \citep{vehtari2017practical} and conformal Bayesian computation \citep{fong2021conformal}. A possible option for estimation is to sample from $\pi(\thetab | \boldsymbol{y}_{1:S})$ (e.g., with \gls{MCMC}). Yet, even i.i.d. sampling from $\pi$ can be highly inefficient in this context, since the variance of $ \widehat{I}_{\text{MC}}^{N} $ takes an interpretable form as $\propto ~ \chi^2(  \pi(\thetab | \boldsymbol{y}_{1:S+1}) ||  \pi(\thetab | \boldsymbol{y}_{1:S}))$. Outliers or influential points lead to large divergences \citep{peng1995bayesian}, so this variance can be unacceptably large. We may instead turn to \gls{SNIS} with a proposal $q(\thetab)$ \citep{vehtari2017practical}, 
whose asymptotic variance is instead $\propto ~~ \chi^2(  \pi(\thetab | \boldsymbol{y}_{1:S+1}) || q(\thetab) ) + \chi^2(  \pi(\thetab | \boldsymbol{y}_{1:S}) || q(\thetab) ) - 2 \cdot \mathcal{C}(q(\thetab)) $ (as we derive in \Cref{sec:main_methodology}). This expression clearly suggests us to choose a $q$ that is close to the posteriors $\pi(\thetab | \boldsymbol{y}_{1:S+1})$ and $\pi(\thetab | \boldsymbol{y}_{1:S})$, and that also maximizes the third term, $\mathcal{C}(q(\thetab))$. For realistic posterior distributions, finding such a $q$ is challenging, which motivates the use of two proposal methods \citep{rainforth2020target}. However, these methods can be seen as taking into account only the first two terms of the asymptotic variance.
\end{example}

\section{Generalized self-normalized importance sampling}\label{sec:main_methodology}
   Next, we derive a decomposition of the asymptotic variance that provides new, clear intuition about how the proposal should be chosen when using \gls{SNIS} and explains its limitations. In \Cref{sec:jointprop_intro}, we introduce our new framework, where samples are obtained from a joint distribution in an extended space. We propose definitions of optimal joint distributions and a two-stage optimization strategy to combine any existing \gls{AIS} method to adapt the marginals of the joint. In \Cref{sec:construction_joint}, we show how expressive joint distributions, possibly semi-parametric, with marginals of different parametric families can be used. Then, in \Cref{sec:optimalcoupling} and \Cref{sec:paramcoupling}, we propose ways to parameterize, sample and optimize the joint. Finally, in \Cref{sec:connecs} we unify some previous methods as special cases, and in \Cref{sec:theory} we develop theoretical guarantees for our generalized SNIS.

\textbf{A motivating variance decomposition.} 
We briefly recall the definition of the \gls{SNIS} estimator from \Cref{sec:intro}. Given \gls{i.i.d.} samples $\boldsymbol{x}^{(n)} \sim q$, the \gls{SNIS} estimator is given as a ratio of two \gls{UIS} estimators,
\begin{equation}
    \SNISest = \frac{\frac{1}{N}\sum_{n=1}^N w_n \cdot \varphi(\boldsymbol{x}^{(n)})}{\frac{1}{N}\sum_{n=1}^N w_n}  = \frac{\hatj}{\hatz} , \quad w_n = \frac{\widetilde{\pi}(\boldsymbol{x}^{(n)})}{q(\boldsymbol{x}^{(n)})} .
\end{equation}
The following decomposition of the asymptotic variance of the \gls{SNIS} estimator lies behind the reasoning for our proposed methodology. It is a rearrangement of the classical delta-method expression \citep{mcbook} which more clearly elucidates how the proposal should be chosen. To introduce it, recall from \Cref{sec:intro} that $\SNISest$ is a ratio of two \gls{UIS} estimators.
\begin{proposition}\label{prop:snis-asym-var}
    Let $\varphi\in L^1(\pi)$ and $H=\mathbb{E}_\pi[|\varphi|]>0$. The variance-minimizing \gls{UIS} proposals for $J$ and $Z_\pi$ are $q_{1,\text{UIS}}^{\bigstar}(\boldsymbol{x})=|\varphi(\boldsymbol{x})|\pi(\boldsymbol{x})/H$ and $q_{2,\text{UIS}}^{\bigstar}(\boldsymbol{x})=\pi(\boldsymbol{x})$, abbreviated to $q_1^{\bigstar}$ and $q_2^{\bigstar}$. If $q_i^{\bigstar}\ll q$ and $q_i^{\bigstar}/q\in L^2(q)$ for $i=1,2$, the asymptotic variance is
    \begingroup\small
    \begin{equation}\label{eq_main_snis}
    \begin{aligned}
        \mathbb{V}_{q}^{\infty}[\SNISest]
        ={}& H^2\{1+\chi^2(q_1^{\bigstar}\|q)\}+I^2\{1+\chi^2(q_2^{\bigstar}\|q)\} -2IH\,\overbrace{\mathbb{E}_{q}\left[\operatorname{sign}\{\varphi(\boldsymbol{x})\}\frac{q_1^{\bigstar}(\boldsymbol{x})q_2^{\bigstar}(\boldsymbol{x})}{q(\boldsymbol{x})^2}\right]}^{:=\mathcal C(q)}.
    \end{aligned}
    \end{equation}
\endgroup
    Here $\chi^2(p\|q)=\int_{\mathcal X}\{p(\boldsymbol{x})/q(\boldsymbol{x})-1\}^2q(\boldsymbol{x})\,d\boldsymbol{x}$ is the Pearson chi-squared divergence, and $\mathcal C(q)$ is a cross moment term. For $\varphi\geq0$, $H=I$ and $\operatorname{sign}\{\varphi(\boldsymbol{x})\}$ can be omitted.
\end{proposition}
We stress that the numerator's optimal proposal $q_{1}^{\bigstar}$ depends on $\varphi$, which need not be bounded. \Cref{eq_main_snis} shows how the variance depends both on how closely $q$ matches the two optimal densities, and on the dependence between the weights. The second-moment assumptions make the three terms well defined, though the ratio can have finite asymptotic variance without them, as we will discuss in \Cref{sec:theory}.

  \Cref{eq_main_snis} suggest a simple story for how to choose the proposal. The first two terms force $q$ to be ``close'' (in the $\chi^2$-sense) to the two optimal \gls{UIS} proposals $q_{1}^{\bigstar},q_{2}^{\bigstar}$. The third term $\mathcal{C}(q)$, deriving from $\operatorname{Cov}_{q}[\hatj,\hatz]$ (see Appendix~\ref{app:deriv}), can affect the asymptotic variance significantly (see \Cref{sec:theory}). While previous works in ratio estimation do consider the use of two proposals $q_1,q_2$ to (implicitly) minimize the first two terms, as discussed in \Cref{sec:intro}, they do not take $\mathcal{C}(q)$ into consideration in the construction of $q$.\footnote{In \Cref{sec:connecs}, we review these works showing their estimators and $\SNISest$ are special cases of our generic framework introduced next.}

\subsection{Introducing an extended space joint proposal}\label{sec:jointprop_intro}
 We now construct a generalized version of the \gls{SNIS} estimator by obtaining samples from a joint distribution in an extended space, $\mathbb{Q}_{1:2}(d\boldsymbol{x}_1,d\boldsymbol{x}_2) \in \mathbb{P}(\mathcal{X} \times \mathcal{X} )$ with marginals $\mathbb{Q}_{1}(d\boldsymbol{x}_1),\mathbb{Q}_{2}(d\boldsymbol{x}_2)$,

\begin{align}\label{eq_gensnis_basic}
      \GenSNISest \eqdef \frac{\hatj}{\hatz} = \frac{\frac{1}{N} \sum_{n=1}^{N} \frac{\varphi(\boldsymbol{x}_{1}^{(n)})  \widetilde{\pi}(\boldsymbol{x}_{1}^{(n)})}{q_{1}(\boldsymbol{x}_{1}^{(n)})}}{\frac{1}{N} \sum_{n=1}^{N} \frac{\widetilde{\pi}(\boldsymbol{x}^{(n)}_{2})}{q_{2}(\boldsymbol{x}^{(n)}_{2})}} , ~~ 
    \begin{array}{l}
    [\boldsymbol{x}_{1}^{(n)},\boldsymbol{x}^{(n)}_{2}]^\top  \iidsim\mathbb{Q}_{1:2}(d\boldsymbol{x}_1,d\boldsymbol{x}_2) , \\  n=1,\dots,N . 
    \end{array} 
\end{align}
For the two component averages to estimate $J$ and $Z_\pi$, respectively, we require $|\varphi|\widetilde\pi\,d\boldsymbol{x}\ll q_1\,d\boldsymbol{x}$ and $\widetilde\pi\,d\boldsymbol{x}\ll q_2\,d\boldsymbol{x}$. The zero-denominator convention mentioned earlier applies here when all $q_2$ weights become zero.

The key element of our new methodology is the use of a generic joint distribution $\mathbb{Q}_{1:2}$, and its construction (which we detail in \Cref{sec:construction_joint}). Marginals of $\mathbb{Q}_{1:2}$ need to admit densities $q_1$ and $q_2$ to compute the estimator. Yet, since the joint $\mathbb{Q}_{1:2}$ does not appear in the expression of $\GenSNISest $, we do not require its density to exist. Exploiting this, we will formalize a semi-parametric model for the joint, which allows us to interpret $\SNISest$ formally as a special case, as well as other methods using two proposals (see \Cref{sec:connecs} for a detailed discussion of the existing literature). Extended spaces are commonplace in advanced \gls{MC} methods, notably  in annealed \gls{IS} and \gls{SMC} samplers \citep{del2006sequential}. We here provide yet another way to exploit them.

Next, we turn to the key question of how to design $\mathbb{Q}_{1:2}$. We extend the asymptotic variance decomposition of \Cref{eq_main_snis} to $\GenSNISest$ to obtain an objective function for the definition of an optimal joint in terms of asymptotic variance.
\begin{proposition}[Asymptotic variance of $\GenSNISest$]\label{thm:asym_var} Let $\varphi\in L^1(\pi)$ and $H=\mathbb{E}_\pi[|\varphi|]>0$, with $q_1^{\bigstar}=|\varphi|\pi/H$ and $q_2^{\bigstar}=\pi$ as above. Assume $q_i^{\bigstar}\ll q_i$ and $q_i^{\bigstar}/q_i\in L^2(q_i)$ for $i=1,2$. For i.i.d. pairs from $\jointprop$, the asymptotic variance of $\GenSNISest$ satisfies (proof in the Appendix~\ref{app:deriv})
    \begin{equation}\label{eq_main_decomp}
    \resizebox{0.92\linewidth}{!}{$\displaystyle
        \mathbb{V}_{\jointprop}^{\infty}[\GenSNISest]
        = H^2\{1+\chi^2(q_1^{\bigstar}\|q_1)\}
        +I^2\{1+\chi^2(q_2^{\bigstar}\|q_2)\}
        -2IH\,\overbrace{\mathbb{E}_{\jointprop}\left[
        \operatorname{sign}\{\varphi(\boldsymbol{x}_1)\}
        \frac{q_1^{\bigstar}(\boldsymbol{x}_1)q_2^{\bigstar}(\boldsymbol{x}_2)}
        {q_1(\boldsymbol{x}_1)q_2(\boldsymbol{x}_2)}
        \right]}^{\mathcal C(\jointprop)}.
    $}
    \end{equation}
    When $\varphi\geq0$, $H=I$ and the sign factor can be omitted so the expression is simpler.
\end{proposition}
The decomposition in \Cref{eq_main_decomp} naturally suggests the following strategy: we seek a joint such that its marginal $q_1$ minimizes the first divergence $\chi^2(q_1^{\bigstar}\|q_1)$ and its marginal $q_2$ minimizes the second divergence $\chi^2(q_2^{\bigstar}\|q_2)$. With these being fixed, we can aim for further variance reduction by adapting a joint which optimizes the third term. Since this term appers with coefficient $-2IH$, increasing it reduces the asymptotic variance when $I>0$, while decreasing it reduces the asymptotic variance when $I<0$. The sign must be known or estimated with a pilot run, but this is not a problem in practice: the sign is difficult to estimate only when $I$ is very close to zero, in which case (i) either it is a rare events problem with $\varphi \geq 0$ (ii) or an accurate estimate with many digits of precision is not needed. 
\begin{remark}[A general framework for combining two adaptive IS or \gls{VI} algorithms]
    By separating the adaptation of the marginals $q_1$, $q_2$ from that of the full joint $\jointprop$, we allow a new framework for adaptive IS algorithms where a first run uses an essentially arbitrary adaptive IS algorithm (or \gls{VI}), to obtain $q_1,q_2$.\footnote{In fact, we can use any algorithm for learning $q$ that implicitly or explicitly minimizes a chi-squared divergence, as long as it returns a \gls{PDF}.} Then, one optimizes the joint to minimize asymptotic variance by maximizing $\operatorname{sign}(I)\mathcal{C}(\jointprop)$. We will show in \Cref{sec:construction_joint} that expressive families of joints and marginals can be used in practice within the framework, such as mixtures of Gaussians, or Student-t distributions. Our framework therefore naturally generalizes and extends \citep{rainforth2020target}, who only propose to learn the marginals (in our interpretation). While \citep{rainforth2020target} also empirically tried annealed \gls{IS} - a special case of sequential MC samplers, themselves closely connected to adaptive IS \citep{portier2018asymptotic} - for numerator and denominator, their theoretical analysis restricts to i.i.d. samples, which is also our setup for theoretical analysis.
\end{remark}

\textbf{Towards defining an optimal joint.} The natural next step is to attempt to formally define a suitable notion of optimal joint $\jointprop$ with fixed marginals. To do so, we define a suitable set of joint distributions for searching over. 

 We define the set of all joint distributions with fixed marginals $\mathbb{Q}_{1} \in \proposalspaceone, \mathbb{Q}_{2} \in \proposalspacetwo$, as \begin{align}\label{eq:semiparametric}
    \Pi(\proposalspaceone,\proposalspacetwo)
    \eqdef \left\{\jointprop\in\mathbb{P}(\mathcal{X}\times\mathcal{X})\ \middle|\ 
    \begin{aligned}
    \int_{\boldsymbol{x}_2}\jointprop(d\boldsymbol{x}_{1:2})&=\mathbb{Q}_1,\\
    \int_{\boldsymbol{x}_1}\jointprop(d\boldsymbol{x}_{1:2})&=\mathbb{Q}_2
    \end{aligned}\right\},
    \end{align}    
where $\boldsymbol{x}_{1:2} = [\boldsymbol{x}_1,\boldsymbol{x}_2]^\top$. We can then define an (asymptotically) optimal joint as follows.

\begin{definition}[Asymptotically optimal joint for $\GenSNISest$ for two given marginals]\label{def:optimaljoint}
     Under the assumptions of \Cref{thm:asym_var}, let $q_1$ and $q_2$ be two given marginal densities. We define an asymptotically optimal joint proposal $\jointprop^{\bigstar}(d\boldsymbol{x}_{1:2})$ for $\GenSNISest$ as any solution of the following infinite-dimensional constrained optimization problem:
    \begin{align}\label{eq:opt_joint}
        \jointprop^{\bigstar} \in \argmax_{\jointprop\in\Pi(\proposalspaceone,\proposalspacetwo)}
        \operatorname{sign}(I)\,
        \mathbb{E}_{\jointprop}\left[
        \operatorname{sign}\{\varphi(\boldsymbol{x}_1)\}
        \frac{q_1^{\bigstar}(\boldsymbol{x}_1)q_2^{\bigstar}(\boldsymbol{x}_2)}
        {q_1(\boldsymbol{x}_1)q_2(\boldsymbol{x}_2)}
        \right].
    \end{align}
\end{definition}

If $q_1=q_1^{\bigstar}$ and $q_2=q_2^{\bigstar}$, the denominator weight is constant and the asymptotic variance is $H^2-I^2$ for every coupling. This variance is zero only when $\varphi$ has a constant sign. Theoretical analyses of previous two proposal methods like \citep{rainforth2020target} focus on cases where marginals are optimal, $q_1 = q_1^{\bigstar}$ and $q_2 = q_2^{\bigstar}$. Then it may be that little additional variance reduction is possible by optimizing the joint (and no further reduction is needed for $\varphi$ with constant sign). However, it is irrealistic to have exactly $q_1 = q_1^{\bigstar}$ and $q_2 = q_2^{\bigstar}$ in practical problems. 

\begin{remark}[An optimization objective without second moments]\label{rem:broader-optimal-joint}
The variance decomposition leading to the objective \Cref{eq:opt_joint} separates the second moments of the two importance weights $q_{i}^{\bigstar}/q_i$ from their cross moment, requiring these to be finite. As we discuss more in detail in \Cref{sec:theory}, studying behaviour of IS often requires one to deal with weights of infinite second moments. Let us introduce a more general criterion than \Cref{eq:opt_joint} which applies in these settings. Based on the observation that the ratio error $\GenSNISest - I =(Z_{\pi}/\hatz)N^{-1} \sum_{n=1}^N H\operatorname{sign}\{\varphi(\boldsymbol{x}_1)\}
\frac{q_1^{\bigstar}(\boldsymbol{x}_1)}{q_1(\boldsymbol{x}_1)}
-
I\frac{q_2^{\bigstar}(\boldsymbol{x}_2)}{q_2(\boldsymbol{x}_2)}$, we can directly define 
\begin{equation}\label{eq:broader}
\argmin_{\jointprop\in\Pi(\proposalspaceone,\proposalspacetwo)}
\mathbb{E}_{\jointprop}\left[
\left\{
H\operatorname{sign}\{\varphi(\boldsymbol{x}_1)\}
\frac{q_1^{\bigstar}(\boldsymbol{x}_1)}{q_1(\boldsymbol{x}_1)}
-
I\frac{q_2^{\bigstar}(\boldsymbol{x}_2)}{q_2(\boldsymbol{x}_2)}
\right\}^{2}
\right],
\end{equation}
provided this expectation is finite for at least one joint. When the separate second moments are finite, expanding the square recovers \Cref{eq:opt_joint}. The objective also takes an interesting optimal transport form, with a typical squared difference cost - except that the (here, weighted for a signed $\varphi$) squared difference is not on the samples $\boldsymbol x_1, \boldsymbol x_2$ directly but on nonlinear transformations of them.

The above objective however less directly leads to a straightforward optimization scheme. Unlike in \Cref{eq:opt_joint}, where unknown multiplicative constants can be dropped for the purposes of optimization, in \cref{eq:broader} they appear multiplying the IS weights inside the difference. A concrete method would need plug-in estimates of these unknowns.
\end{remark}

 Next, we will define concrete constructions for the joint distribution so that the optimization problem can be addressed. In particular, we will focus on marginals that are in parametric families $\proposalspaceoneparametrizedplain, \proposalspacetwoparametrizedplain$. The methodological challenge then is to choose/learn the joint parameters while keeping the marginals fixed.
\begin{remark}
        Choosing typical families used in \gls{IS} like \glspl{MVN} imposes restrictions on marginal families. For instance, if we were to consider $\jointprop(d\boldsymbol{x}_1,d\boldsymbol{x}_2; \thetabplain)$ to be \gls{MVN} with $\thetabplain = [\muplain, \Sigmaplain]$, i.e., with moment parameters $\bmu = [\muplain_1,\muplain_2] \in \mathbb{R}^{2 \dimens}, \boldsymbol{\Sigma} = \left [ \begin{smallmatrix}
        \Sigmaplain_1 & \boldsymbol{\Sigma}_{1 \times 2}^\top \\ 
        \boldsymbol{\Sigma}_{1 \times 2} & \Sigmaplain_2
        \end{smallmatrix} \right ] \in \mathcal{S}_{++}^{2 \dimens \times 2 \dimens} $, then $\mathbb{Q}_{1}$ and $\mathbb{Q}_{2}$ are restricted to be \glspl{MVN} by definition. The same restriction applies to the multivariate Student-t family. 
        The need for two different parametric families can easily arise in practice. Consider our motivating example from \Cref{mot:example}: if test data is sampled from a Student-t model, with the posterior being \gls{MVN}, then the two marginals of our joint proposal should be heavy-tailed (Student) and \gls{MVN}.     
\end{remark}

Motivated by these limitations, we next develop a general method to sample from a joint distribution with two given marginals of \textbf{(i)} possibly different parametric families, and \textbf{(ii)} such that the parametric family (of the joint) is not in a known or named family. 

\subsection{Constructing the joint with couplings and Brenier maps}\label{sec:construction_joint}
Our proposed construction of the joint $\jointprop$ decomposes it into a coupling (defined next) and its marginals. We also note how the construction naturally allows for a certain type of normalizing flow (NF) marginals. NFs are a natural choice for complex high-dimensional \gls{IS}, and have been shown to have the potential for competitive performance even with SMC samplers in some high-dimensional examples \citep{andrade2024stabilizing}. See \hyperref[app:normalizing-flows]{the Appendix} for a discussion on how to incorporate NFs.
We start with a useful definition. 

\begin{definition}[Coupling]\label{def:coupling}
     A coupling is a joint probability distribution $\couplingmeasure$ for random vectors $\boldsymbol{u}_1,\boldsymbol{u}_2 \in [0,1]^{\dimens}$ such that the marginals of $\boldsymbol{u}_1$ and $\boldsymbol{u}_2$ are uniform, i.e., $\boldsymbol{u}_1 \sim \mathbb{U}_{[0,1]^{\dimens}}(d \boldsymbol{u}_1), \boldsymbol{u}_2 \sim \mathbb{U}_{[0,1]^{\dimens}}(d \boldsymbol{u}_2)$. We denote its density, when it exists, as $c_{1:2}(\boldsymbol{u}_1,\boldsymbol{u}_2)$.
\end{definition}
 For $\dimens = 1$, the above definition amounts to a copula \citep{czado2019analyzing}. The notion of copula is not sufficient for our purposes: it cannot model multivariate marginals \citep{genest1995impossibilite}.

 \textbf{Sampling from a joint with fixed marginals.} We now discuss how to obtain samples from a joint with given multivariate marginals, a nontrivial task in general \citep{kazi2019construction}. Assume that \textbf{(i)} we know how to sample from a coupling $\couplingmeasure$ and \textbf{(ii)} have access to measurable functions $\boldsymbol{T}_i : [0,1]^{\dimens} \rightarrow \mathbb{R}^{\dimens}$ that transport the uniform distribution to the desired marginals, i.e., $\boldsymbol{T}_i \sharp \mathbb{U}_{[0,1]^{\dimens}}(d \boldsymbol{u}_i) = \mathbb{Q}_i(d \boldsymbol{x}_i)$ for $i=1,2$. Then, we can obtain a sample $[\boldsymbol{x}_1,\boldsymbol{x}_2]^\top$ with the following two steps.
 \begin{itemize}
    \item \textbf{Step 1 (Coupling sampling)}: Obtain a sample $[\boldsymbol{u}_1,\boldsymbol{u}_2] \sim \couplingmeasure$.
    \item \textbf{Step 2 (Marginal transformation)}: Apply the maps $\boldsymbol{T}_i$ to $\boldsymbol{u}_i$ to obtain $\boldsymbol{x}_i = \boldsymbol{T}_i(\boldsymbol{u}_i)$, for $i=1,2$. Denote the joint distribution of $\boldsymbol{x}_1,\boldsymbol{x}_2$ as $\jointprop(d \boldsymbol{x}_1, d \boldsymbol{x}_2)$. 
 \end{itemize}
The maps $\boldsymbol{T}_i$ hold the marginals fixed while the coupling determines how the samples are paired. We now detail the two steps.   \textbf{Step 1 (Coupling sampling).} As a possible way to implement step (1), we could exploit the use of an easy-to-deal-with auxiliary joint distribution $\mathbb{G}_{1:2}(d \boldsymbol{x}_{1}^\prime, d \boldsymbol{x}_{2}^\prime)$, e.g., a \gls{MVN} or a Student-t distribution, by sampling from it $[\boldsymbol{x}_{1}^{\prime},\boldsymbol{x}_{2}^{\prime}] \sim \mathbb{G}_{1:2}(d \boldsymbol{x}_{1}^\prime, d \boldsymbol{x}_{2}^\prime) $ and transforming the sample into one from a coupling by applying the inverse maps $\boldsymbol{T}_{i}^{-1}: \mathbb{R}^{\dimens} \rightarrow [0,1]^{\dimens}$ to transform the marginals $ \boldsymbol{T}_{i}^{-1} \sharp \mathbb{G}_{i}(d \boldsymbol{x}_i) = \mathbb{U}_{[0,1]^{\dimens}}(d \boldsymbol{u}_i)$ for $i=1,2$ as before. Concretely, for $\dimens = 1$, $T_{i}$ can be chosen as the inverse \glspl{CDF} (quantile) of the marginals. In general, for $\dimens > 1$, we propose the use of \glspl{VQ} introduced in \citep{10.1214/16-AOS1450}. Interestingly, \glspl{VQ} coincide exactly the celebrated optimal transport Brenier maps, as noted in \citep{carlier2020vector}.\footnote{We note that our coupling sampling method is a generalization of the method described in \citep[Chapter 5, Section 6]{mcbook}.}   \textbf{Step 2 (marginal transformation).}
  Brenier maps are defined as any $\boldsymbol{T}_i \sharp \mathbb{U}_{[0,1]^{\dimens}}(d \boldsymbol{u}_i) = \mathbb{Q}(d \boldsymbol{x}_i)$ that can be expressed as the gradient of a convex function, i.e., there exists a convex $ \psi(\boldsymbol{u}_i): [0,1]^{\dimens} \rightarrow \mathbb{R} \cup \{ + \infty \}$ such that $  \boldsymbol{T}_i \eqdef \nabla_{\boldsymbol{u}_i} \psi(\boldsymbol{u}_i) $. While the definition is abstract, \citep{fan2023vector} shows that they can be constructively found for interesting distributions commonly used in \gls{IS}, e.g., for \gls{MVN}, Student-t marginals, and more. When these maps are diffeomorphisms, their Jacobians give the marginal densities required by the estimator.

This construction has an important property: it can represent every joint distribution with the chosen marginals. Indeed, under the conditions of the vector version of Sklar's theorem, each joint has a unique representing coupling once the marginal Brenier maps are fixed \citep[Theorem~1]{fan2023vector}. Thus searching over all couplings is equivalent to searching over all joints with these marginals. We give details in the Appendix~\ref{app:joint-generality}.
 
 In \Cref{fig:illustration_couplings}, we give an illustration of the proposed process of sampling from a joint with fixed marginals. 
 We stress that in general, to compute the estimator $\GenSNISest$, the coupling (hence, the joint distribution) does not itself require a density. Indeed, some relevant couplings that we incorporate do not admit a density, including the implicit one used by \gls{SNIS} (see \Cref{sec:connecs}). We expand on this in \Cref{sec:connecs}.
\begin{figure}[!t]
    \centering
    \begin{subfigure}[t]{0.3\linewidth}
        \centering
        \includegraphics[width=\linewidth,height=0.2\textheight,keepaspectratio]{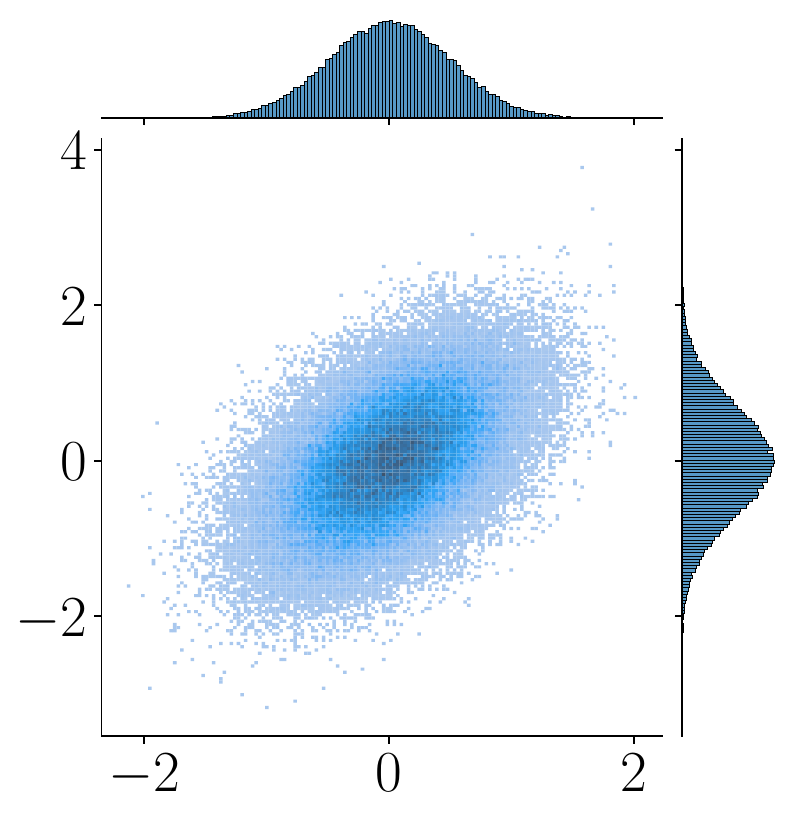}
        \caption{Gaussian joint.}
        \label{fig:mvn}
    \end{subfigure}%
    \hfill
    \begin{subfigure}[t]{0.3\linewidth}
        \centering
        \includegraphics[width=\linewidth,height=0.2\textheight,keepaspectratio]{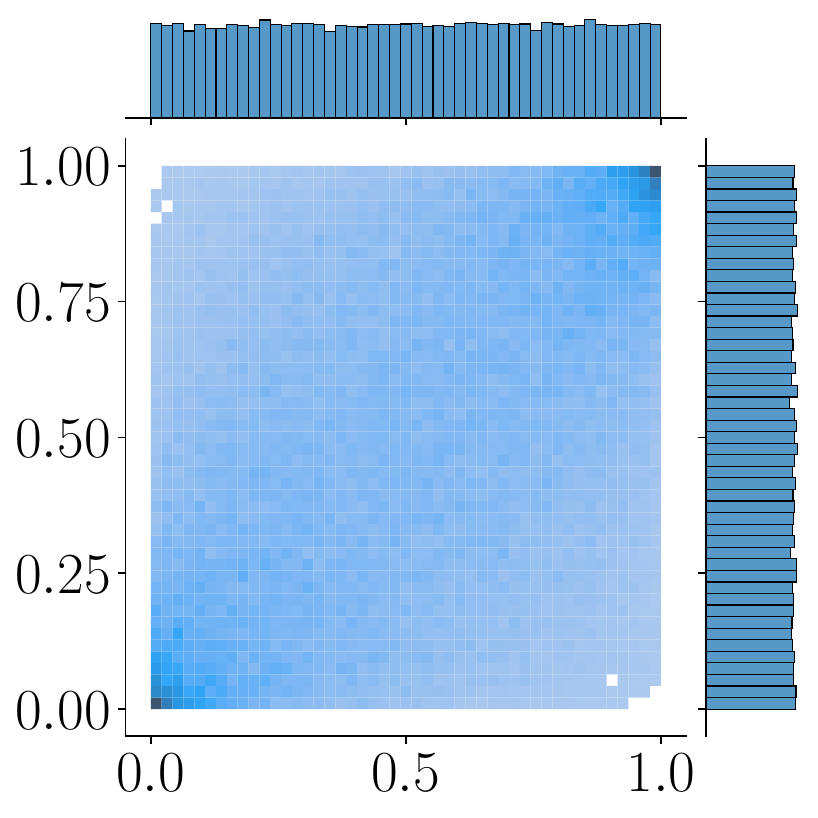}
        \caption{Uniform marginals.}
        \label{fig:mvn_coupling}
    \end{subfigure}%
    \hfill
    \begin{subfigure}[t]{0.3\linewidth}
        \centering
        \includegraphics[width=\linewidth,height=0.2\textheight,keepaspectratio]{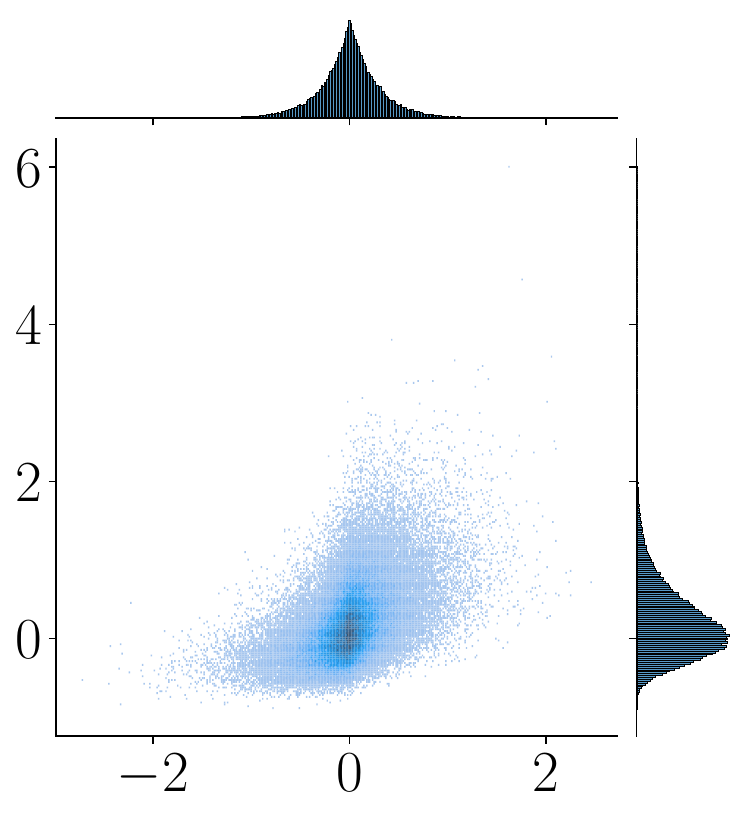}
        \caption{Laplace--Gumbel marginals.}
        \label{fig:joint_mvn_laplace_gumbel}
    \end{subfigure}

    \begin{subfigure}[t]{0.3\linewidth}
        \centering
        \includegraphics[width=\linewidth,height=0.2\textheight,keepaspectratio]{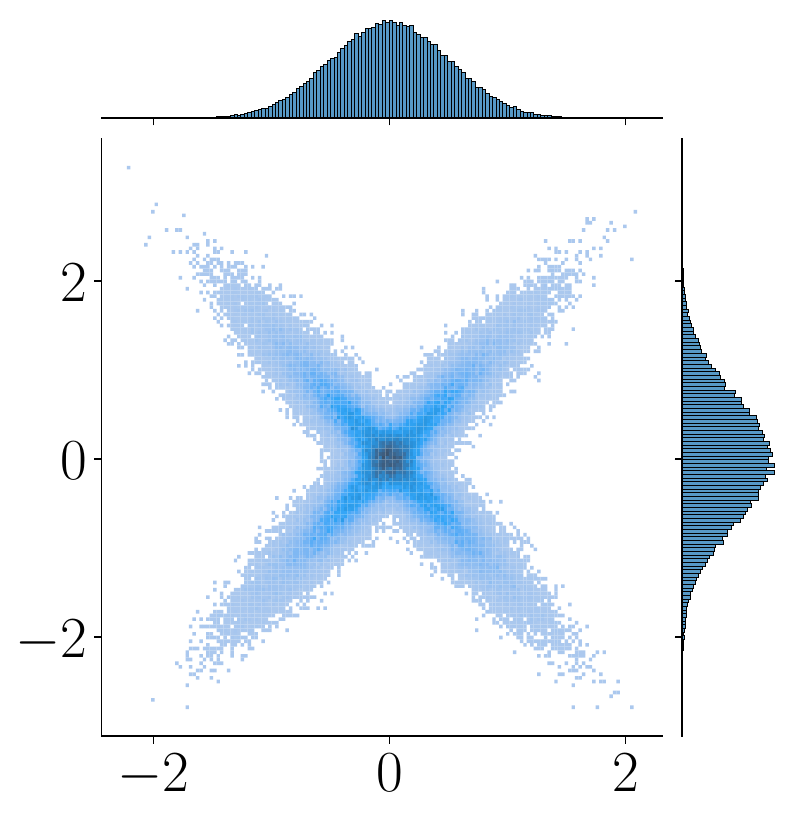}
        \caption{Gaussian mixture.}
        \label{fig:gauss_mix}
    \end{subfigure}%
    \hfill
    \begin{subfigure}[t]{0.3\linewidth}
        \centering
        \includegraphics[width=\linewidth,height=0.2\textheight,keepaspectratio]{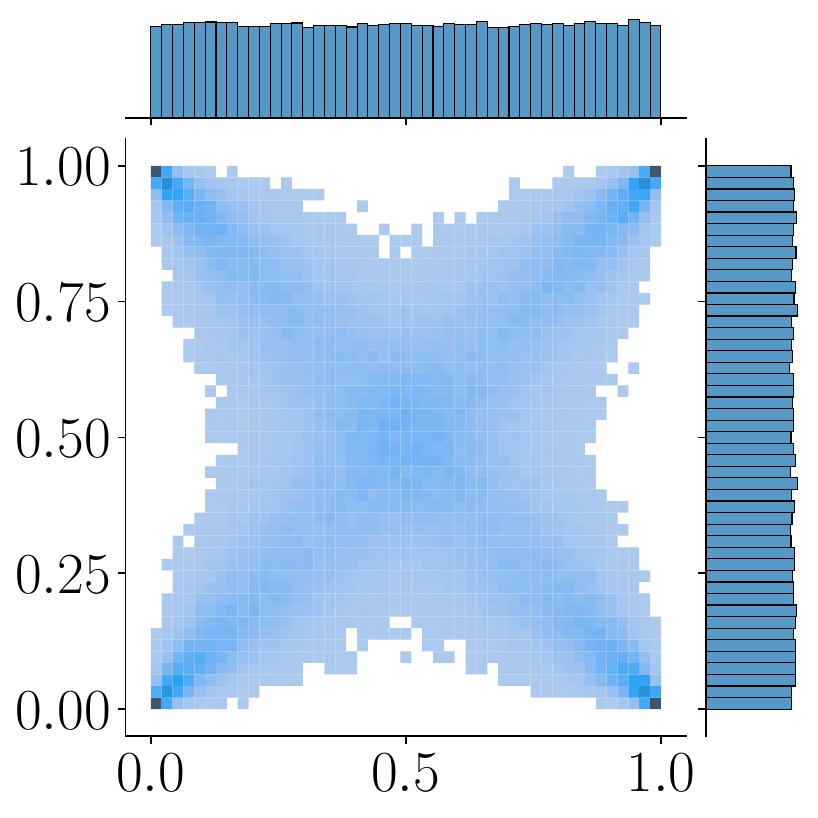}
        \caption{Uniform marginals.}
        \label{fig:gauss_mix_coupling}
    \end{subfigure}%
    \hfill
    \begin{subfigure}[t]{0.3\linewidth}
        \centering
        \includegraphics[width=\linewidth,height=0.2\textheight,keepaspectratio]{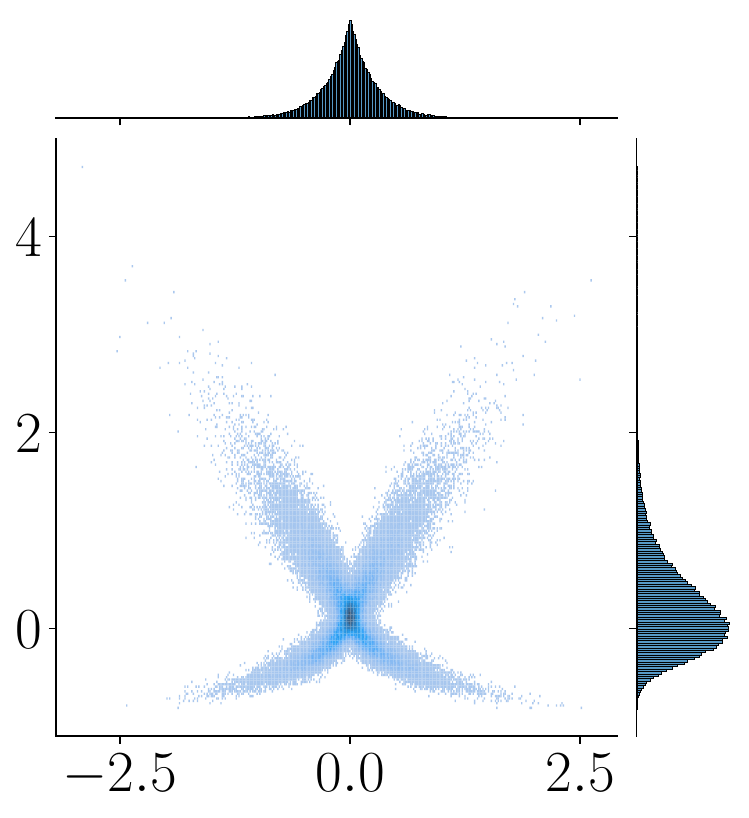}
        \caption{Laplace--Gumbel marginals.}
    \end{subfigure}
    \caption{\textbf{Flexibility of our $\jointprop$ construction.} The first column shows a Gaussian joint with correlation $0.5$ (top) and a mixture with correlations $-0.9$ and $0.9$ (bottom), both with standard Gaussian marginals. Applying the marginal CDFs gives the couplings in the middle column; applying Laplace and Gumbel quantile functions gives the last column. Within each column the marginals agree, while each row preserves its coupling.}
    \label{fig:illustration_couplings}
\end{figure}

\subsection{From optimal joint to optimal coupling}\label{sec:optimalcoupling}
We are now equipped to rewrite \Cref{eq:opt_joint} directly in terms of the coupling $\mathbb{C}_{1:2}$, which will conveniently turn a constrained optimization problem into an unconstrained one, under our general construction.

Rewriting the objective in \Cref{eq:opt_joint} by applying the change of variables $\boldsymbol{x}_i \rightarrow \mathbf{T}_{i}^{-1}(\boldsymbol{x}_i) = \boldsymbol{u}_i$ leads to 
\begin{align}\label{eq:to_maximize}
\mathcal C(\jointprop)
&=\mathbb{E}_{\jointprop}\!\left[
\operatorname{sign}\{\varphi(\boldsymbol{x}_1)\}
\frac{q_1^{\bigstar}(\boldsymbol{x}_1)q_2^{\bigstar}(\boldsymbol{x}_2)}
{q_1(\boldsymbol{x}_1)q_2(\boldsymbol{x}_2)}\right]
=\mathbb{E}_{\mathbb C_{1:2}}[\omega(\boldsymbol{u}_1,\boldsymbol{u}_2)].
\end{align}
where the term inside the expectation is a kind of joint-space importance weight in the hypercube $[0,1]^{2 \dimens}$, given by
\begin{equation}\label{eq:newoptimization}
\omega(\boldsymbol{u}_1,\boldsymbol{u}_2)
=\operatorname{sign}\{\varphi(\boldsymbol{T}_1(\boldsymbol{u}_1))\}
\omega_1(\boldsymbol{u}_1)\omega_2(\boldsymbol{u}_2),\quad
\omega_i(\boldsymbol{u}_i)
=\frac{q_i^{\bigstar}(\boldsymbol{T}_i(\boldsymbol{u}_i))}
{q_i(\boldsymbol{T}_i(\boldsymbol{u}_i))},\quad i=1,2.
\end{equation}

Using the same change of variable, we can also express the broader criterion in \Cref{rem:broader-optimal-joint}, which allows infinite separate second moments for numerator and denominator, as
$$
\mathbb{E}_{\mathbb C_{1:2}}\!\left[
\left\{H\operatorname{sign}\{\varphi(\boldsymbol{T}_1(\boldsymbol{u}_1))\}
\omega_1(\boldsymbol{u}_1)-I\omega_2(\boldsymbol{u}_2)\right\}^2\right].
$$

\subsection{Concrete choices of coupling}\label{sec:paramcoupling}

Next, we discuss concrete coupling choices that can be used in our framework. We will also discuss a method for coupling adaptation to maximize \Cref{eq:opt_joint}, i.e., explicitly targeting an asymptotically optimal coupling $\mathbb{C}_{1:2}^{\bigstar}$. We stress, however, that it is \emph{not} our purpose here to propose the best possible method for learning a coupling - our emphasis in on the flexibility that our framework introduces allowing for new methods to be developed.

\textbf{Multivariate Normal and Student-t couplings.}
We start considering the coupling that arises as a consequence of selecting the auxiliary distribution described in \Cref{sec:construction_joint} as \gls{MVN}. A \gls{MVN} coupling $\mathbb{C}_{1:2}^{\texttt{MVN}}(d\boldsymbol{u}_1, d\boldsymbol{u}_2; \boldsymbol{S}_c)$ is the distribution of $\boldsymbol{u}_1,\boldsymbol{u}_2$ obtained by sampling $[\z_1,\z_2] \iidsim \mathcal{N}(\boldsymbol{0}, \boldsymbol{\Sigma}_c ) $  where $\boldsymbol{\Sigma}_c = \left [ \begin{smallmatrix} 
    \boldsymbol{I}_{\dimens} & \boldsymbol{S}_{c}^{\top} \\ 
    \boldsymbol{S}_{c} & \boldsymbol{I}_{\dimens } 
    \end{smallmatrix} \right ] $and setting $\boldsymbol{u}_{i} = \Phi(\z_i)$, where $\Phi$ is the \gls{CDF} of the standard normal and is applied element-wise. In particular, to sample from $\mathcal{N}(\boldsymbol{0}, \boldsymbol{\Sigma}_c )$ one can use a matrix square root $\boldsymbol{L} :  \boldsymbol{L}\boldsymbol{L}^\top = \boldsymbol{\Sigma}_c$, which will have the structure $\boldsymbol{L} = \left [\begin{smallmatrix} 
    \boldsymbol{I}_{\dimens} & \boldsymbol{0} \\ 
    \boldsymbol{S}_{c} & \boldsymbol{M} 
    \end{smallmatrix} \right ]$ where $\boldsymbol{M}$ is a matrix such that $\boldsymbol{M}\boldsymbol{M}^\top = \boldsymbol{I}_{\dimens} - \boldsymbol{S}_{c}\boldsymbol{S}_{c}^\top$. Again, it is worth keeping in mind that using $\mathbb{C}_{1:2}^{\texttt{MVN}}(d \boldsymbol{u}_1, d \boldsymbol{u}_2; \boldsymbol{S}_c)$ does not mean restricting $q_1,q_2$ be \glspl{MVN}. Extending to a Student-t coupling is straightforward. \\ \textbf{Mixture \& neural couplings.}
A mixture of couplings is a valid coupling, leading to straightforward extensions of the \gls{MVN}/Student-$t$ couplings (Appendix~\ref{app:uncon}).\footnote{Differentiation of mixture parameters can use deterministic mixture sampling techniques \citep[Chapter~9]{mcbook}.} To obtain even more flexibility and introduce a coupling exploting neural networks, we consider a simple \emph{continuous} mixture coupling where the cross-covariance between numerator and denominator samples is parameterized. Let $\boldsymbol{G},\boldsymbol{H}\sim\mathcal N_{\dimens}(\boldsymbol{0},\boldsymbol{I}_{\dimens})$ and $\boldsymbol{\eta}\sim\mathcal N_r(\boldsymbol{0},\boldsymbol{I}_r)$ be independent, let $\boldsymbol{U},\boldsymbol{V}$ be orthogonal, and let $\boldsymbol{\alpha}_{\boldsymbol{\psi}}(\boldsymbol{\eta})\in(-1,1)^{\dimens}$. Then, set
$$
\boldsymbol{Z}=\boldsymbol{G},\qquad
\boldsymbol{Z}'=\boldsymbol{U}\left\{\boldsymbol{\alpha}_{\boldsymbol{\psi}}(\boldsymbol{\eta})\odot \boldsymbol{V}^{\mathsf T}\boldsymbol{G}
+\sqrt{\boldsymbol{1}-\boldsymbol{\alpha}_{\boldsymbol{\psi}}(\boldsymbol{\eta})^2}\odot \boldsymbol{H}\right\}.
$$
 Conditional on $\boldsymbol{\eta}$, both inputs are standard Gaussian and $\operatorname{Cov}(\boldsymbol{Z}',\boldsymbol{Z}\mid\boldsymbol{\eta})=\boldsymbol{U}\operatorname{diag}\{\boldsymbol{\alpha}_{\boldsymbol{\psi}}(\boldsymbol{\eta})\}\boldsymbol{V}^{\mathsf T}$. Thus, $\Phi(\boldsymbol{Z})$ and $\Phi(\boldsymbol{Z}')$ have uniform marginals which, as before, can be transformed to the desired ones. To learn the nonlinear mapping $\boldsymbol{\alpha}_{\boldsymbol{\psi}}$, we can use a neural network with input $\boldsymbol{\eta}$ and output $\boldsymbol{\alpha}_{\boldsymbol{\psi}}(\boldsymbol{\eta})$.
We describe a further neural coupling based on rotations of the Gaussian inputs in Appendix~\ref{app:uncon}. Also in the Appendix, we describe how we can incorporate some simple couplings used in the variance reduction literature (in other contexts than the SNIS estimator) as initializations for our more complex optimized couplings. 

\subsection{Special cases falling within our framework}\label{sec:connecs}
We now show, more formally, that the estimator $\SNISest$ is special case of $\GenSNISest$, formalizing the implicit assumption of using the same samples to estimate the integrals $J$ and $Z_{\pi}$ as discussed in \Cref{sec:intro}. We also show how several existing methods can be viewed as special cases, and that the flexibility of our joint construction allows one to combine existing couplings with more generic marginal families in a principled way.

 \textbf{Couplings based on common random numbers.} Within our framework, sharing the same samples in both integrals, as done by $\SNISest$, corresponds to choosing a joint singular probability measure  $\mathbb{Q}^{\text{SNIS}}_{1:2}(d\boldsymbol{x}_1,d\boldsymbol{x}_2) = \delta_{\boldsymbol{x}_1}(d \boldsymbol{x}_2) \cdot \mathbb{Q}_{1}(d \boldsymbol{x}_1) = \delta_{\boldsymbol{x}_2}(d \boldsymbol{x}_1 ) \cdot \mathbb{Q}_{2}(d \boldsymbol{x}_2)$  where \textbf{(i)} for almost every $\boldsymbol{x}_1$ (respectively, $\boldsymbol{x}_2$) the regular conditional distribution of one vector given the other is a Dirac measure and \textbf{(ii)} the marginals are equal, $ \mathbb{Q}_{1}=\mathbb{Q}_{2}$.\footnote{A singular probability measure does not admit a density w.r.t. the Lebesgue measure.} The corresponding coupling is then also a singular probability measure; by a change of variable, 
\begin{equation}\label{eq:crncoupling}
    \mathbb{C}_{1:2}^{\text{CRN}}(d\boldsymbol{u}_1,d\boldsymbol{u}_2)
    =\boldsymbol{T}_{1:2}\sharp\mathbb{Q}_{1:2}^{\text{SNIS}}(d\boldsymbol{x}_1,d\boldsymbol{x}_2)
    =\delta_{\boldsymbol{u}_1}(d\boldsymbol{u}_2)\cdot\mathbb{U}_{[0,1]^{\dimens}}(d\boldsymbol{u}_1),
\end{equation}
where \gls{CRN} stands for ``common random numbers'' and we use the abstract vector quantile transformations described earlier  $\boldsymbol{T}_{1:2} \eqdef [\boldsymbol{T}_1(d \boldsymbol{x}_1), \boldsymbol{T}_2(d \boldsymbol{x}_2)]^\top$ applied to the product space of measurable subsets of $\mathbb{R}^{\dimens}$. \cite{fan2023vector} notice that this coupling is a generalization of the so-called comonotonic copula \citep[Chapter 5]{mcbook}. Similarly, we can of course have $\delta_{\boldsymbol{1} - \boldsymbol{u}_1}(d \boldsymbol{u}_2) \cdot \mathbb{U}_{[0,1]^{\dimens}}(d \boldsymbol{u}_1)$ as another coupling, corresponding to the counter-monotonic copula.\footnote{Mixing the two would be possible, setting some components as $u_{1,1} = u_{2,1}$ and others as $u_{1,3} = 1 - u_{2,3}$.} The above formalizes and generalizes the method to sample from $\mathbb{C}_{1:2}^{\text{CRN}}$ (and variants) which we mentioned in \Cref{sec:paramcoupling} and describe fully in the Appendix.

\textbf{Independent coupling.} When we use the same number of samples for numerator and denominator and independent proposals, we recover the two-proposal estimator studied by \citet[Section~2.1, Eq.~(2.1)]{chen1997monte}, as well as the more recent target-aware Bayesian inference (TABI) \citep{rainforth2020target} (see also \citep{llorente2023target}). Their corresponding coupling is
\begin{align}\label{eq:indep_coupling}
    \mathbb{C}_{1:2}^{\text{INDEP}}(d \boldsymbol{u}_1, d \boldsymbol{u}_2) = \mathbb{U}_{[0,1]^{\dimens}}(d \boldsymbol{u}_1) \cdot \mathbb{U}_{[0,1]^{\dimens}}(d \boldsymbol{u}_2) . 
\end{align}
Crucially, this coupling cannot exploit cancellation due to numerator/denominator dependence for variance reduction. Two different marginals were also used by \citep{lamberti2018double}. They linearly transform a Gaussian random variable $x_1 \in \mathbb{R}$ into $x_2 \in \mathbb{R}$, with their method not extending easily to $\dimens > 1$ and specific to Gaussian marginals. The latter, within our interpretation, samples from a base distribution $q$ and transforms the samples to have two different marginals $q_1,q_2$, also implicitly using the \gls{CRN} coupling.\footnote{It is worth noting that a paper proposing the use of two different distributions $q_1,q_2$ for numerator and denominator in the context of \gls{IS} first appeared in \citep{goyal1987measure}.} These previous works do not study the influence of the coupling on the asymptotic variance. A full overview of the methods falling within our framework is given in \Cref{table:specialcases}. Our unifying framework, along with the interpretable variance decomposition in \Cref{eq_main_decomp} allows us to study which choices of couplings can be appropriate in which settings, and paves the way for improving over them.

\begin{table}[!htb]
    \small
    \caption{We show how our framework generalizes many previous methods in the literature. $\mathbb{C}_{1:2}^{\text{CRN}}$ is defined in \Cref{eq:crncoupling}, $\mathbb{C}_{1:2}^{\text{INDEP}}$ is defined in \Cref{eq:indep_coupling}. }
\begin{tabularx}{\columnwidth}{YYY}
\toprule
\textbf{Special case of $\GenSNISest$} & \textbf{Choice of marginals} & \textbf{Choice of coupling} \\
\midrule
        \gls{SNIS}  &  $q_1(\boldsymbol{x}_1) = q_2(\boldsymbol{x}_2)$  & $\mathbb{C}_{1:2} = \mathbb{C}_{1:2}^{\text{CRN}}$ \\[10pt]
        \gls{UIS}  & Any $q_{1}(\boldsymbol{x}_1)$, $q_2(\boldsymbol{x}_2) = p(\boldsymbol{x}_2)$ & $\mathbb{C}_{1:2}  = \mathbb{C}_{1:2}^{\text{CRN}}$ \\[10pt]
        \citep{lamberti2018double}  & $q_1(x_1) \neq q_2(x_2)$ \color{black} & $\mathbb{C}_{1:2} = \mathbb{C}_{1:2}^{\text{CRN}}$ \\[10pt]
         \citep{goyal1987measure}; \citep{chen1997monte}; \citep{golinski2018,rainforth2020target} & Arbitrary $q_1,q_2$, possibly equal & $\mathbb{C}_{1:2} =  \mathbb{C}_{1:2}^{\text{INDEP}} $ \\[10pt]
        Equal marginals, dependent (new)  & $q_1(\boldsymbol{x}_1) = q_2(\boldsymbol{x}_2)$ & Any $\mathbb{C}_{1:2}$ \\[10pt]

        \bottomrule
    \end{tabularx}
    \label{table:specialcases}
\end{table}

\subsection{Theoretical properties of generalized SNIS}\label{sec:theory}

We start by briefly reviewing the most relevant theoretical results from the \gls{IS} literature to our work. We re-emphasize that our main focus is on optimizing performance for a specific test function $\varphi$, rather than across a class of test functions. This is in contrast to much of the \gls{IS} literature (although it is common in rare event problems), which focuses on uniform error bounds over classes of test functions. For example, \citet[Theorem~2.1]{agapiou2017importance} provide uniform mean squared error bounds for \gls{SNIS} over bounded functions, and their Theorem~2.3 extends this to unbounded functions under additional moment assumptions. \citet[Theorems~2.1 and~2.2]{deligiannidis2024importance} derive bias and moment bounds under conditions on the weights and on the test function. In contrast, we are fundamentally interested in a test function-specific approach, following our motivating examples and the recent literature in Bayesian statistics addressing these problems \citep{rainforth2020target,paananen2021implicitly}. \cite{sanz2018importance} has sample size guidelines for IS, but they do not cover SNIS. \cite{chatterjee2018sample} has the most related results, which we will discuss in detail.

Throughout this Section, we will provide several insights into the theoretical properties of our generalized \gls{SNIS} estimator. We begin with an easily overlooked point in the analysis of SNIS: that the asymptotic variance can be finite even when both numerator and denominator have infinite variance. In this case, the specific decomposition of the asymptotic variance in \Cref{eq_main_decomp} does not apply anymore. Yet, by inspecting the correct condition for finite asymptotic variance of the whole ratio, we will notice an interesting role played by the coupling: we can take an independent numerator and denominator pair, each having infinite variance, and show that there exists a coupling that reduces the asymptotic variance to a finite value. This insight leads us to then analyse the key quantity playing a role in the asymptotic variance further, and derive a new sufficient sample size bound for the generalized \gls{SNIS} estimator inspired by \citep{chatterjee2018sample}, which explicitly shows the role of the coupling. 

 Throughout, we deal with a general $\varphi\in L^1(\pi)$ and $H=\mathbb{E}_\pi|\varphi|>0$, and recall $q_1^{\bigstar}=|\varphi|\pi/H$ and $q_2^{\bigstar}=\pi$ from \Cref{eq_main_snis}, also their support assumptions. We will also use repeatedly the key quantity 
 \begin{align}\label{eq:res_key}
 r(\boldsymbol{x}_1,\boldsymbol{x}_2) = H\operatorname{sign}\{\varphi(\boldsymbol{x}_1)\} \frac{q_1^{\bigstar}(\boldsymbol{x}_1)}{q_1(\boldsymbol{x}_1)} - I \frac{q_2^{\bigstar}(\boldsymbol{x}_2)}{q_2(\boldsymbol{x}_2)} ,
 \end{align}
which we can recall from \Cref{rem:broader-optimal-joint} in the main text, and recall that the asymptotic variance of SNIS is the second moment of $r$. 
\paragraph{Cancellation through a suitable coupling: from infinite to finite asymptotic variance.}
Throghout the paper, we discussed the benefit of allowing flexibility to have two proposals, $q_1 \neq q_2$, which was exploited in previous methods discussed in \Cref{sec:intro} and \Cref{sec:connecs}. In real problems, it is common for importance weights to be unbounded \citep{agapiou2017importance,deligiannidis2024importance}. In our perspective, we have seen that the SNIS estimator is really a ratio of two UIS estimators, which themselves are both averages of the two weights $q_i^{\bigstar}/q_i$. Of course, unbounded random variables can have finite variance, but it is common in realistic problems to have infinite variance. Then, it is natural in such situations to wonder whether the asymptotic variance of the overall estimator $\GenSNISest$ can still be finite nonetheless.

\textbf{Observation (Finite asymptotic variance by coupling).}
There exist a target density $\pi$, a test function $\varphi\in L^1(\pi)$ with $I\neq0$, and marginals $q_1,q_2$ satisfying the support and positivity mentioned earlier, for which
$
\mathbb E_{q_i}\left[\left(\frac{q_i^{\bigstar}}{q_i}\right)^2\right]
=\infty,i=1,2,
$
but that admit a joint proposal $\jointprop$ with these marginals such that it holds
$
\mathbb E_{\jointprop}[r^2]<\infty.
$
On the other hand, a factorised $\jointprop = \mathbb Q_1\otimes\mathbb Q_2$ with these marginals gives $\mathbb E_{\mathbb Q_1\otimes\mathbb Q_2}[r^2]=\infty$. Our observation comes from a simple but typically overlooked fact within \gls{SNIS}: finite variance of the whole estimator only requires second moments of the residual $r$, not individual moments of numerator and denominator. We think this provides strong motivation for pursuing new choices of couplings within the generalized SNIS framework and going beyond two proposal methods of previous works.

\paragraph{Chatterjee-Diaconis type bounds for generalized SNIS.}
We have discussed that a suitable coupling can bring the \emph{asymptotic} variance down from infinite to finite, carefully looking at the role of the residual error $r(\boldsymbol x_1,\boldsymbol x_2)$. However, \cite{chatterjee2018sample} demonstrated that asymptotic variance can be an overly strict criterion for good performance of an estimator. Meaning, that we can have an estimator being close almost all of the time to the true value, even if its variance is infinite, typically due to a few outliers. They then proposed non-asymptotic bounds on the probability of making large $L_1$ errors which do not assume finite variance of the estimators, including \gls{SNIS}. They use these bounds to provide both sufficient and necessary sample size guidance for good performance in IS. However, their results in Theorem 1.2 about \gls{SNIS}: (i) do not accurately take into account $\varphi$; (ii) do not elucidate the effect of the numerator/denominator dependence and (iii) assume bounded $||\varphi||_{L_{2}(\pi)} < \infty$, making their bound uninformative for functions $\varphi \in L_{1}(\pi) \setminus L_{2}(\pi)$. In fact, given the focus on $L_1$ errors for the estimator, we find it natural to seek guarantees for $\varphi \in L_{1}(\pi) \setminus L_{2}(\pi)$ as well.

\textbf{Tilting $\jointprop$ by the residual error.} Our analysis will be based on a key distribution defined as
\begin{align}\label{eq:joint_target}
\mathbb Q_r(d\boldsymbol{x}_1,d\boldsymbol{x}_2)
:=\frac{|r(\boldsymbol{x}_1,\boldsymbol{x}_2)|}{\mathbb{E}_{\jointprop}|r|}\,\jointprop(d\boldsymbol{x}_1,d\boldsymbol{x}_2).
\end{align}
 This distribution concentrates on samples $\boldsymbol{x}_1,\boldsymbol{x}_2$ that contribute most to the error. Both $r$ and of course $\jointprop$ here depend on the coupling choice. Note that for ordinary SNIS (which corresponds to a specific choice of $\jointprop$), then $\mathbb Q_r$'s marginals reduce to the optimal proposal for SNIS \citep{hesterberg1988advances}, discussed in \Cref{sec:intro}. Now we present our probabilistic bounds on the $L_1$ error for generalized SNIS (and hence SNIS), which unlike \citep{chatterjee2018sample} satisfies (i)-(iii) mentioned above. This makes it also the first such non-asymptotic result explicitly involving the optimal proposal for SNIS. The $L_1$ error is the absolute value of error that we analysed earlier in the paper, $ |\GenSNISest-I| = |(Z_{\pi}/\hatz) \overline{r}_N |$.

 \textbf{Upper bound and sufficient sample size.}
\begin{theorem}[Upper bound for probability of large $L_1$ errors.]\label{thm:coupled-residual-bound}
Suppose $\mathbb{E}_{\jointprop}|r|>0$ and define $L_r<\infty$ as the Kullback-Leibler divergence between $\mathbb Q_r$ and $\jointprop$, $L_r := \operatorname{KL}(\mathbb Q_r || \jointprop)$. Then, letting $t\geq0$, $\epsilon>0$, and $N\in\mathbb N$ with $N\geq\exp(L_r+t)$, the estimator $\GenSNISest$ satisfies 
\begin{equation}\label{eq:coupled-residual-bound}
\resizebox{0.88\linewidth}{!}{$\displaystyle
\mathbb P_{\jointprop^{\otimes N}}\left(\left|\GenSNISest-I\right|>\epsilon\right)
\leq
\mathbb P_{\jointprop^{\otimes N}}(\hatz<Z_\pi/2)
+\frac{2\mathbb{E}_{\jointprop}|r|}{\epsilon}
\left[
e^{-t/4}
+2\mathbb Q_r\left(
\log\frac{|r|}{\mathbb{E}_{\jointprop}|r|}>L_r+\frac t2
\right)
\right].
$}
\end{equation}
 When $I\ne0$, replacing $\epsilon$ by $|I|\epsilon$ gives the corresponding relative error bound.
\end{theorem}
The proof is in Appendix~\ref{app:proof_ub}. The first term controls the probability of a small denominator, which can amplify even small differences in the error $ |\GenSNISest-I| = |(Z_{\pi}/\hatz) \overline{r}_N |$. We followed a similar approach to \citep{deligiannidis2024importance} for this. The second term connects directly with the upper bounds of \cite{chatterjee2018sample}; rather than log IS weights, we have log absolute residual, and the KL divergence is between our joint $\jointprop$ and the newly introduced $\mathbb{Q}_r$ from \Cref{eq:joint_target}. As \cite{chatterjee2018sample}, we now use the above upper bound to give a sample-size guarantee for a chosen accuracy.

 Fix an error tolerance \(\epsilon>0\) and a failure probability \(\delta\in(0,1)\). As \(t\) increases, the residual term in \Cref{eq:coupled-residual-bound} tends to zero, so we can choose \(t\) to make it at most \(\delta/2\). Recalling that $q_2$ is fixed, the probability that \(\hatz<Z_\pi/2\) is at most \(\kappa^N\) for some \(\kappa<1\). Therefore, taking 
\[
N\ge\max\left\{e^{L_r+t},\frac{\log(2/\delta)}{-\log\kappa}\right\}
\]
ensures \(\mathbb P(|\GenSNISest-I|>\epsilon)\le\delta\). The choice of $t$ and the denominator bound are detailed in the Appendix~\ref{app:proof_ub}. 

\textbf{Lower bound and necessary sample size.}
We now consider when fewer samples fail to achieve the desired accuracy for the same $\varphi$ and $\jointprop$. We obtain a lower bound on the probability of error by considering underestimation, $\GenSNISest<I-\epsilon$, which implies an absolute error larger than $\epsilon$.

\begin{theorem}[Lower bound for probability of large $L_1$ errors]\label{thm:coupled-residual-lower-bound}
Assume $0<\mathbb{E}_{\jointprop}|r|<\infty$, $L_r<\infty$, and $\widetilde\pi(\boldsymbol{x}_2)/q_2(\boldsymbol{x}_2)>0$ $q_2$-almost surely. For $\epsilon>0$, $t\geq0$ and every $N\leq e^{L_r-t}$,
\begin{equation}\label{eq:coupled-residual-lower-bound}
\begin{aligned}
&\mathbb P_{\jointprop^{\otimes N}}\bigl(|\GenSNISest-I|>\epsilon\bigr)
\geq1-e^{-t/2}\\
&\quad-\frac{2}{\mathbb{E}_{\jointprop}|r|}
\mathbb{E}_{\jointprop}\!\left[
\left(r+\epsilon\frac{\widetilde\pi(\boldsymbol{x}_2)}{Z_\pi q_2(\boldsymbol{x}_2)}
+\frac{\mathbb{E}_{\jointprop}|r|}{2}\right)_+
\mathbf1_{\{|r|\leq e^{L_r-t/2}\mathbb{E}_{\jointprop}|r|\}}
\right],
\end{aligned}
\end{equation}
where $u_+=\max(u,0)$. See proof in Appendix~\ref{app:residual-lower-bound}.
\end{theorem}
Below the sample size \(N\leq e^{L_r-t}\), all samples from the joint satisfy \(|r|\leq e^{L_r-t/2}\mathbb{E}_{\jointprop}|r|\) with probability at least \(1-e^{-t/2}\). To conclude that $\GenSNISest$ is likely to underestimate \(I\) by more than \(\epsilon\), we must also control the probability that \(\widehat I_{\mathrm{GenSNIS}}^N\geq I-\epsilon\) when all samples satisfy this inequality: this is what the final term controls. When that term is small, missing the large values of \(|r|\) is therefore likely to produce an error larger than \(\epsilon\).%

 We now obtain necessary sample size guarantees exploiting \Cref{thm:coupled-residual-lower-bound}. As before, for a desired failure probability $\delta\in(0,1)$, suppose that some $t\geq0$ makes the final term in \Cref{eq:coupled-residual-lower-bound} strictly smaller than $1-\delta-e^{-t/2}$. Then
$$
\mathbb P_{\jointprop^{\otimes N}}\bigl(|\GenSNISest-I|>\epsilon\bigr)>\delta
\quad\text{for every integer }1\leq N\leq e^{L_r-t}.
$$
Therefore achieving failure probability at most $\delta$ requires $N>e^{L_r-t}$. 
Let us conclude now with further comments in how our results compare with the SNIS result of \citep[Theorem 1.2]{chatterjee2018sample}. As mentioned earlier, our focus is on guarantees for a given function $\varphi$. Their upper bounds and thus sufficient sample size results assume $|| \varphi ||_{L_2(\pi)} <\infty$ and are not, in some sense, for a specific $\varphi$: two $\varphi$'s with the same $|| \varphi ||_{L_2(\pi)}$ obtain the same bound. Their lower bound and thus necessary sample size results show that too small sample sizes cannot guarantee accurate estimation for \emph{every} test function. Our lower bound and necessary sample size results show that, for a given $\varphi$, the sample size must exceed our given threshold for accurate estimation. Our approach and proof techniques followed from the observation that previous approaches to non-asymptotic bounds for SNIS \citep{chatterjee2018sample} (but see also the Appendix of \citep{thin2021neo}) ignore the cross terms when bounding the residual error, thus getting rid of the effect of dependence.

\section{Applications}\label{sec:experiments}
In our experiments, we will demonstrate that a careful control of the dependence between numerator and denominator estimates in $\GenSNISest$ (therefore, with any of its special cases) can lead to significantly better estimates of $\mu$, demonstrating the benefits of our framework. Throughout, we will compare with recent state-of-the-art methods specialised for our task, including the TABI framework \citep{rainforth2020target} as a representative framework for two proposal methods.

 Within our framework, we will use $q_1,q_2$ as marginals of a joint proposal $\mathbb Q_{1:2}$. Then, as discussed in \Cref{table:specialcases}, we have special cases: (i) SNIS with proposal $q_1$; (ii) SNIS with proposal $q_{2}$; (iii) SNIS with a certain choice of proposal ``in-between'' $q_1,q_2$, e.g., a mixture (which we discuss later); (iv) two independent proposals $q_1,q_2$ (equivalent to TABI \citep{rainforth2020target}). Against these baselines, we compare $\GenSNISest$ with the same $q_1,q_2$ marginals, and three possible choices of couplings: (i) GenSNIS with CRN coupling; (ii) GenSNIS with antithetic CRN coupling; (iii) GenSNIS with an optimized/learned parametric coupling. Importantly, for all methods except SNIS, for fair comparison, a total budget of $S$ samples is split into $S / 2$ for use in numerator and denominator. For (iii), we will not take into account in the final estimation the computation used to learn the coupling. Our aim here is to assess our hypothesis throughout the paper that a better joint proposal leads to better estimation in generalized SNIS, while proposing some concrete ways to select or learn this joint. Efficient ways to do so will be the subject of future work. 

\paragraph{Bayesian case deletion/outlier analysis/cross validation with synthetic data.}
Going back to our \Cref{mot:example}, we will test our hypotheses aforementioned in examples where the estimation task of interest is with $\pi$ being a Bayesian posterior distribution, and $\varphi$ being the likelihood of one or more datapoints. This happens in several closely related applications, including in Bayesian cross validation \citep{vehtari2002bayesian,vehtari2017practical}; in the Bayesian analysis of outlier observations, related to so called case-deletion \citep{peng1995bayesian,peruggia1997variability}. What is common in all these applications is the fact that the quantity of interest measures in some sense how influential one or more observations are to the inference, and, as explained in \Cref{mot:example}, the computation of these quantities can be especially challenging when compared to more ``standard'' posterior expectations (such as means, variances). As mentioned earlier, these types of expectations continue to arise in applications, as also put forward by TABI \citep{rainforth2020target} and its application to modern probabilistic programming systems \citep{reichelt2022expectation}. Broadly, quantifying the influence of datapoints to Bayesian inferences continues to gather interest \citep{nguyen2024sensitivity}. 

For our example, we shall consider simulated Bayesian linear regression with $y_i\mid\boldsymbol{\beta}\sim\mathcal N(\boldsymbol{x}_i^{\mathsf T}\boldsymbol{\beta},\sigma^2)$ and $\boldsymbol{\beta}\sim\mathcal N_d(\boldsymbol{0},\tau_d^2\boldsymbol{I}_d)$. We let $A$ be an index set containing $80$ observations out of a total of 100, with $B$ denoting that total. Then, with $\pi(\boldsymbol{\beta})=p(\boldsymbol{\beta}\mid \boldsymbol{y}_A,\boldsymbol{X})$ and $\varphi(\boldsymbol{\beta})=p(\boldsymbol{y}_{B\setminus A}\mid\boldsymbol{\beta},\boldsymbol{X})$, our quantity of interest in \Cref{eq_snis_ratio} becomes $I=\mathbb{E}_\pi[\varphi(\boldsymbol{\beta})]=J/Z_\pi=p(\boldsymbol{y}_{B\setminus A}\mid \boldsymbol{y}_A,\boldsymbol{X})$, where $J=p(\boldsymbol{y}_B\mid \boldsymbol{X})$ and $Z_\pi=p(\boldsymbol{y}_A\mid \boldsymbol{X})$.

Letting $D\in\{A,B\}$, write $\pi_D=\mathcal N_d(\boldsymbol{\mu}_D,\boldsymbol{\Sigma}_D)$. We will select our marginal proposals for all algorithms as perturbed versions of the optimal zero variance ones (which here are available): $q_1=q_B=\mathcal N_d(\boldsymbol{\mu}_B+\boldsymbol{\delta}_B,\boldsymbol{\Sigma}_B)$ and $q_2=q_A=\mathcal N_d(\boldsymbol{\mu}_A+\boldsymbol{\delta}_A,\boldsymbol{\Sigma}_A)$, where $\boldsymbol{L}_D\boldsymbol{L}_D^{\mathsf T}=\boldsymbol{\Sigma}_D$ and $(\boldsymbol{\delta}_D)_j=0.2(\boldsymbol{L}_D)_{jj}\varepsilon_{D,j}/\sqrt d$ for randomly chosen signs $\varepsilon_{D,j}\in\{-1,1\}$. We chose the perturbation to decrease with dimension so that IS errors (at least individually for numerator and denominator) do not explode excessively. We then try SNIS with proposals $q_A,q_B$ and a couple of other choices that we may expect to be better than these two individual ones. The first such option is SNIS with a mixture proposal $m_{AB}=\tfrac12(q_A+q_B)$, even sampling deterministically from each component (instead of randomly) and weighting with the whole mixture (a multiple IS version of SNIS) to reduce variance. The second is another Gaussian (like $q_1,q_2$) whose moments are an average of the moments of $q_1,q_2$, i,e., $
q_{\rm MM}=\mathcal N_d\!\left(
\tfrac12(\bar{\boldsymbol\mu}_A+\bar{\boldsymbol\mu}_B),
\tfrac12(\boldsymbol\Sigma_A+\boldsymbol\Sigma_B)
+\tfrac14(\bar{\boldsymbol\mu}_A-\bar{\boldsymbol\mu}_B)
(\bar{\boldsymbol\mu}_A-\bar{\boldsymbol\mu}_B)^{\mathsf T}
\right).$ Regarding GenSNIS with a learned coupling, we use the continuous mixture of Gaussian couplings in \Cref{sec:paramcoupling}. 
The results in \Cref{fig:blr-dimensions} over $25$ replications show that across all dimensions and both sample budgets in \Cref{fig:blr-dimensions}, learned GenSNIS has the lowest median absolute log-ratio error.

\begin{figure}[!htbp]
    \centering
    \includegraphics[width=0.85\linewidth]{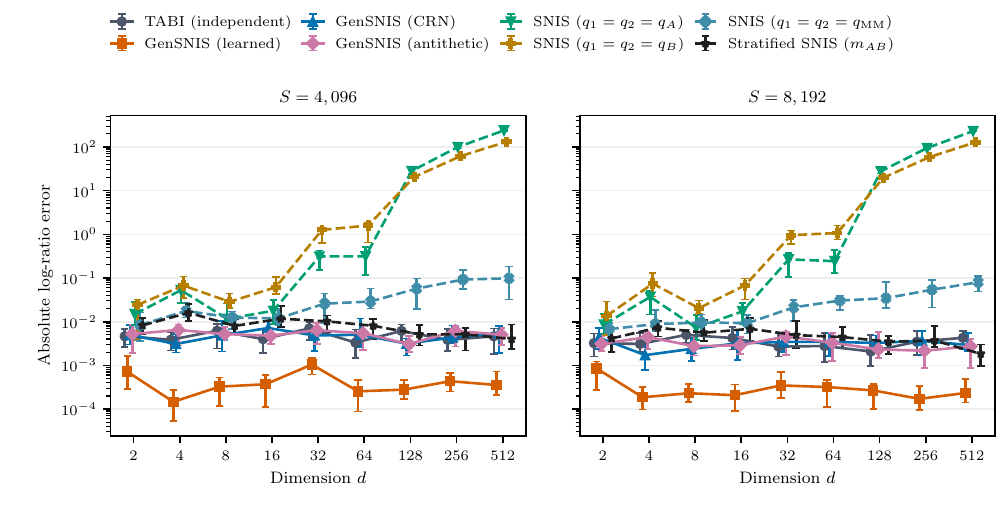}
    \caption{Results for Bayesian linear regression, estimating posterior predictive values. The figure shows the median and interquartile range of $|\log\widehat I-\log I|$ over $25$ repetitions at each dimension $d$. Each panel shows results for total sample budget $S$.}
    \label{fig:blr-dimensions}
\end{figure}

\FloatBarrier

\textbf{Real data application: prior sensitivity in the US Crime data.}

Next, we turn to yet another application where we argue that accurate estimates of individual expectations (i.e., for a specific $\varphi$) are of interest, and it is easy to find realistic examples where SNIS can be significantly outperformed. The task is that of determining how sensitive are our conclusions - in Bayesian inference, these are expectations under the posterior $\pi$ - to certain parameters; these can be prior parameters or others; we focus here on the prior's. This problem has been well studied and goes under various names, including Bayesian robustness or Bayesian sensitivity analysis \citep{giordano2018covariances}. We will show now that the natural quantity of interest for this problem can easily lead to a challenging estimation problem. In particular under commonly used priors, we will obtain an expectation of a $\varphi \in L_{1}(\pi) \setminus  L_{2}(\pi)$ - motivating for careful choice of proposal selection.

We take the 47-state \texttt{UScrime} data, previously used to study prior sensitivity by \citet[Section~5.5]{kallioinen2024detecting}, and analyze them with Gaussian regression. In this dataset the response $y$ is standardized log crime rate; see Appendix~\ref{app:uscrime-coupling} for details on the covariates. Inverse-gamma priors are common for the noise variance $v$ in regression. We will use a beta-prime prior so that we can specify separately how plausible very small and very large noise variances are \citep{perez2017scaled}. As $v$ approaches zero, the beta-prime density decreases polynomially, whereas an inverse-gamma density decreases exponentially. This gives more prior mass to very small values of $v$, allowing for a relationship with little unexplained variation, while retaining a heavy upper tail for larger noise variances. Concretely, the model we use is

$$
\begin{aligned}
\boldsymbol{y}\mid\boldsymbol{\beta},v&\sim\mathcal N_{47}(\boldsymbol{X}\boldsymbol{\beta},v\boldsymbol{I}_{47}),\qquad
\boldsymbol{\beta}\sim\mathcal N_{58}(\boldsymbol{0},\boldsymbol{\Sigma}),\qquad v\sim\operatorname{BP}(5/2,5/2),\\
\boldsymbol{\Sigma}&=0.9\boldsymbol{I}_{58}+0.1\{\boldsymbol{X}^+(\boldsymbol{X}^+)^{\mathsf T}+\boldsymbol{P}_{\ker \boldsymbol{X}}\}.
\end{aligned}
$$
Here $\boldsymbol X^+$ is the Moore--Penrose inverse. To see why the projector $\boldsymbol P_{\ker\boldsymbol X}$ appears, consider a change in $\boldsymbol\beta$ that leaves $\boldsymbol X\boldsymbol\beta$ unchanged. The $\boldsymbol X^+(\boldsymbol X^+)^{\mathsf T}$ term contributes zero variance along that change, while $\boldsymbol P_{\ker\boldsymbol X}$ gives it the $0.1$ component. Since $\boldsymbol X\boldsymbol X^+=\boldsymbol I_{47}$, the prior covariance of $\boldsymbol X\boldsymbol\beta$ is $\boldsymbol X\boldsymbol\Sigma\boldsymbol X^{\mathsf T}=0.9\boldsymbol X\boldsymbol X^{\mathsf T}+0.1\boldsymbol I_{47}$.

To study sensitivity of our conclusions to the prior on $v$, we multiply its median by $e^\alpha$, leaving the rest of the model fixed. Let $\pi_\alpha(\boldsymbol\beta,v\mid\boldsymbol y)$ denote the resulting full posterior. We then consider how the \emph{posterior variance} of $G(\boldsymbol\beta,v)=\boldsymbol\beta^{\mathsf T}\boldsymbol\Sigma^{-1}\boldsymbol\beta/(58v)$ changes with $\alpha$. The derivative of the log prior density at $\alpha=0$ is $s(v)=5/2-5/(1+v)$.

\begin{samepage}
Applying \citet[Theorem~1]{giordano2018covariances} to $G$ and $G^2$ gives
$$
\begin{aligned}
\theta
&=\left.\partial_\alpha\operatorname{Var}_{\pi_\alpha(\boldsymbol\beta,v\mid\boldsymbol y)}(G)\right|_{\alpha=0}
=\mathbb E_{\pi_0(\boldsymbol\beta,v\mid\boldsymbol y)}[\varphi(\boldsymbol\beta,v)],\\
\varphi(\boldsymbol\beta,v)
&=\left\{G-\mathbb E_{\pi_0(\boldsymbol\beta,v\mid\boldsymbol y)}[G]\right\}^2
\left\{s(v)-\mathbb E_{\pi_0(\boldsymbol\beta,v\mid\boldsymbol y)}[s(v)]\right\}.
\end{aligned}
$$
\end{samepage}

To see why this expectation is difficult to estimate, consider small values of $v$. The matrix $\boldsymbol X$ has 58 columns but only 47 rows, so changes in $\boldsymbol\beta$ that leave $\boldsymbol X\boldsymbol\beta$ fixed still vary under the posterior. The factor $1/v$ in $G$ magnifies their contribution. Together with the beta-prime prior's behavior near zero, this gives $\mathbb E_{\pi_0(\boldsymbol\beta,v\mid\boldsymbol y)}[|\varphi|]<\infty$ but $\mathbb E_{\pi_0(\boldsymbol\beta,v\mid\boldsymbol y)}[\varphi^2]=\infty$. This is an integrable test function of the kind considered in \Cref{sec:theory} (see Appendix~\ref{app:uscrime-coupling} for the derivation). Let us now discuss how we chose the proposal marginals $q_1,q_2$. Importantly, we will choose $q_1,q_2$ that, when used in SNIS (which recall, uses only one proposal, and we try both), are expected to be good choices and certainly better than the special case $q = \pi_{0}(\boldsymbol{\beta},v | \boldsymbol{y})$ which has infinite variance. This is so that the improvements we shall get with generalized SNIS are even more informative. 

To choose a sensible $q_1(\boldsymbol\beta,v)$ we need it to put more mass on small values of $v$. Before proceeding, note that in this model, we can write down the conditional posterior of $\boldsymbol\beta$ analytically as $\pi_0(\boldsymbol\beta\mid v,\boldsymbol y)=\mathcal N_{58}(\boldsymbol m_v,\boldsymbol C_v)$. Define $\rho(v)=\{\mathbb E[\varphi^2\mid v,\boldsymbol y]\}^{1/2}$ and $K=\int_0^\infty\rho(v)\pi_0(v\mid\boldsymbol y)\,dv$. Among proposals that retain this Gaussian conditional, a variance marginal proportional to $\pi_0(v\mid\boldsymbol y)\rho(v)$ minimizes the second moment of the numerator importance weight. We use its mixture with the full joint posterior,
$q_1(\boldsymbol\beta,v)=\pi_0(\boldsymbol\beta,v\mid\boldsymbol y)\{0.05+0.95\rho(v)/K\}$ as typical in defensive importance sampling.

Regarding the denominator proposal, we slightly increase the conditional covariance and mix with $q_1$:
$q_2(\boldsymbol\beta,v)=0.95\,\pi_0(v\mid\boldsymbol y)\mathcal N_{58}(\boldsymbol\beta;\boldsymbol m_v,1.10\boldsymbol C_v)+0.05q_1(\boldsymbol\beta,v)$.
This small $q_1$ component also helps SNIS using $q_2$ against the small-$v$ numerator tail. 

Finally, to learn the coupling for generalized SNIS, we couple Gaussian inputs to the two proposals through six conditional rotations. We describe the construction in detail in the Appendix~\ref{app:uscrime-coupling}.

\FloatBarrier
\begin{figure}[H]
    \centering
    \includegraphics[width=\linewidth]{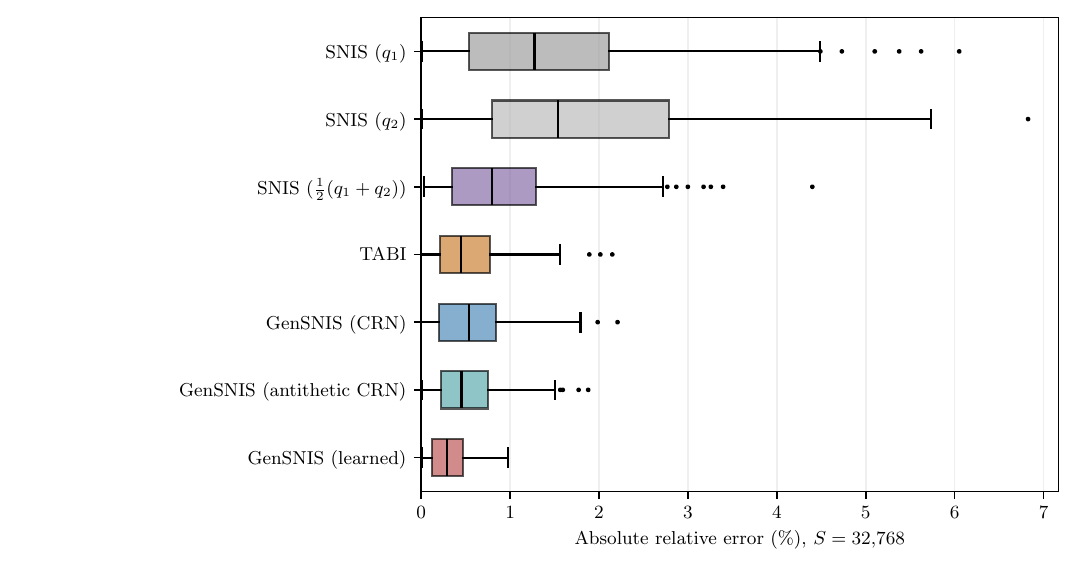}
    \caption{Prior sensitivity in the US Crime data. Boxplots (1.5-IQR, outliers shown) give absolute relative errors $|\widehat\theta-\theta|/|\theta|$ at $S=32{,}768$ total samples over $200$ independent replications.}
    \label{fig:uscrime-sensitivity}
\end{figure}

The learned coupling gives the lowest errors in \Cref{fig:uscrime-sensitivity}. At $S=32{,}768$, its mean absolute relative error is $0.315\%$, against $0.541\%$ for TABI, $0.597\%$ for CRN, and $0.534\%$ for antithetic CRN. SNIS with $\tfrac12(q_1+q_2)$ gives $0.963\%$. With the same fixed proposal marginals, learning reduces the error by $41$--$47\%$ relative to TABI and the two simple couplings, and by $67\%$ relative to mixture SNIS.

\FloatBarrier

\section{Discussion \& conclusions}\label{sec:exp_discussion}
The SNIS estimator is ubiquitous, even beyond Bayesian statistics and machine learning. We have introduced a broad generalization of SNIS, exploiting its view as a ratio estimator. Our framework, generalizing existing methods using two proposals for numerator and denominator, samples from a joint distribution in an extended space. Our joint construction is very general and flexible, accomodatiing different parametric families for the marginals and complex couplings.

We have shown theoretical and empirical benefits resulting from our perspective. Specifically, we have shown that an appropriate choice of coupling can bring the asymptotic variance from infinite to finite. We have then also proved new probability bounds on the $L_1$ error of generalized SNIS (hence, also applying to SNIS), inspired by the results of \citep{chatterjee2018sample}, leading to new necessary and sufficient sample size guarantees when a $\varphi$ is given and specific to SNIS. 

We expect that it will be possible to design more sophisticated joint distributions depending on the application and that doing so can help address difficult estimation problems, where the interest is in estimating accurately expectations for a specific $\varphi$. A promising direction of future work is to propose principled extensions which \emph{amortize} the selection of the joint (both marginals and coupling) so that these are suitable for many expectations at once, which is typical and a complementary focus to our scope here. While existing two proposal methods propose amortized extensions \citep{rainforth2020target}, it is difficult to obtain theoretical guarantees of performance for a class of expectations.

\endgroup

\section*{Acknowledgements}
\AdditionalAcknowledgments
We would like to thank several people for their feedback and useful discussions on this work and some of its earlier versions. In chronological order, we thank Simo Sarkka for interesting discussions on the functional optimization problem for the joint at BayesComp 2023; to Art Owen for generous comments and feedback at MCM 2023; to Christophe Andrieu for valuable suggestions on how the theory could be extended and relevant related work at the SMC workshop in 2024. We again thanks Art Owen for useful feedback on a more recent version of this work and for suggesting to look into inverse moments assumptions to handle small denominators. We thank El Mahdi Khribch for pointing us to his paper whose approach was also innstrumental in several of our results to handle small denominators. 

\section*{Statement on AI use}
We used OpenAI ChatGPT and Codex to assist with coding and running experiments. We also used them to assist with the proof of the necessary sample size. In particular, Codex suggested to simply assume an a.s. positive denominator estimate leading to a simple if and only if condition for underestimation, which then led us to arrive at Eq S2.12. For the sufficient sample size, Codex made more rigorous our argument for bounding the probability of a small denominator. The paper is human written. 
\ifdefempty{\ArxivFundingStatement}{}{%
  \section*{Funding}\ArxivFundingStatement}
\ifdefempty{\ArxivDisclosureStatement}{}{%
  \section*{Disclosure Statement}\ArxivDisclosureStatement}
\ifdefempty{\ArxivDataAvailabilityStatement}{}{%
  \section*{Data Availability Statement}\ArxivDataAvailabilityStatement}

\bigskip
\section*{Appendix}
\pdfbookmark[0]{Appendix}{arxiv-appendix}
\setcounter{section}{0}
\renewcommand{\thesection}{S\arabic{section}}
\renewcommand{\theHsection}{appendix.\arabic{section}}
\renewcommand{\theHsubsection}{\theHsection.\arabic{subsection}}
\renewcommand{\theHsubsubsection}{\theHsubsection.\arabic{subsubsection}}
\glsresetall
\providecommand{\circledd}[1]{\circled{#1}}
\providecommand{\abs}[1]{\lvert#1\rvert}

\section{Concrete coupling constructions and optimization}\label{app:uncon}
In this Section we explain the parameterizations used in Section~\ref{sec:paramcoupling} more in detail. We also describe the conditional rotations used in the US Crime experiment.

\paragraph{Orthogonal factors and diagonal parameters.}
To enforce the constraint $\|\boldsymbol{S}_c\|_{\mathrm{op}}\leq1$, we write $\boldsymbol{S}_c=\boldsymbol{U}\operatorname{diag}(\boldsymbol{\sigma})\boldsymbol{V}^{\mathsf T}$, with orthogonal $\boldsymbol{U},\boldsymbol{V}$ and $\sigma_j\in[-1,1]$. In the BLR implementation, each orthogonal factor is a product of Householder reflections. For example,
$$
\boldsymbol{U}=\boldsymbol{P}(\boldsymbol{a}_1)\cdots\boldsymbol{P}(\boldsymbol{a}_K),
\qquad
\boldsymbol{P}(\boldsymbol{a})=\boldsymbol{I}_{\dimens}
-2\frac{\boldsymbol{a}\boldsymbol{a}^{\mathsf T}}{\boldsymbol{a}^{\mathsf T}\boldsymbol{a}},
\qquad \boldsymbol{a}\ne\boldsymbol{0}.
$$
The matrix $\boldsymbol{P}(\boldsymbol{a})$ reflects a vector across the hyperplane perpendicular to $\boldsymbol{a}$. Since a reflection is orthogonal, we can vary the nonzero vectors $\boldsymbol{a}_1,\ldots,\boldsymbol{a}_K$ freely and the product of the corresponding reflections will remain orthogonal. We use the same construction for $\boldsymbol{V}$. To keep the diagonal entries in $(-1,1)$ during optimization, we set $\sigma_j=(2/\pi)\arctan(v_j)$ and optimize over $v_j\in\mathbb R$. For CRN and full antithetic CRN, we instead set $\boldsymbol{S}_c=\boldsymbol{I}_{\dimens}$ and $\boldsymbol{S}_c=-\boldsymbol{I}_{\dimens}$, respectively.

\paragraph{Conditional rotations.}
For greater flexibility, we can let the angle of a rotation vary with its input. We must choose this dependence carefully so that the transformed input still has the standard Gaussian distribution. Let $\boldsymbol{z}\sim\mathcal N_{\dimens}(\boldsymbol{0},\boldsymbol{I}_{\dimens})$, choose $i\ne j$, and write $\boldsymbol{z}_{-ij}$ for the remaining coordinates. Given a neural network $\gamma_{\boldsymbol{\psi}}$, set
$$
\begin{pmatrix}z_i'\\z_j'\end{pmatrix}
=\boldsymbol{R}_{\gamma_{\boldsymbol{\psi}}(z_i^2+z_j^2,\boldsymbol{z}_{-ij})}
\begin{pmatrix}z_i\\z_j\end{pmatrix},
\qquad \boldsymbol{z}_{-ij}'=\boldsymbol{z}_{-ij},
\qquad
\boldsymbol{R}_\gamma=
\begin{pmatrix}\cos\gamma&-\sin\gamma\\\sin\gamma&\cos\gamma\end{pmatrix}.
$$
Conditional on the radius of $z_j', z_i'$ and the remaining coordinates, the angle is uniform. Under this conditioning, the network output is fixed. Then rotating a uniform angle obviously leaves it uniform. The radius and the other coordinates stay unchanged, so the transformed vector is again standard Gaussian. We can repeat the construction on different pairs of coordinates to obtain $\boldsymbol{C}_{\boldsymbol{\psi}}$. The sample $(\boldsymbol{z},\boldsymbol{C}_{\boldsymbol{\psi}}(\boldsymbol{z}))$ then has standard Gaussian marginals with a singular joint distribution, which is not an issue in our framework, as we do not need a density for the joint. In Section~\ref{app:uscrime-coupling} we describe the coordinates and reconstruction used in the US Crime experiment.

\textbf{Common random numbers as special cases and initialization.} We show how to incorporate in our framework couplings commonly used in the variance reduction literature \citep{mcbook}. We discuss formally in Section~\ref{sec:connecs} their intepretation in terms of the implied joint distribution. These couplings simply set $\boldsymbol{u}_1 = \boldsymbol{u}_2$ or simple variations such as $\boldsymbol{u}_1 = \boldsymbol{1} - \boldsymbol{u}_2$. We now generalize and interpret them as special cases of sampling from the \gls{MVN} coupling. As a consequence, we will be able to \emph{initialise} the parameter $\boldsymbol{S}_c$ of a \gls{MVN} coupling at the value that corresponds to using a \gls{CRN} coupling. To start with, we will represent $\boldsymbol{S}_c$ via its \gls{SVD}, $\boldsymbol{S}_{c} = \boldsymbol{U}_{c} ~ \text{diag}(\sigma_1,\dots,\sigma_{\dimens}) ~\boldsymbol{V}_{c}^\top $ with $\boldsymbol{U}_{c}$ and $\boldsymbol{V}_{c}$ orthogonal matrices and singular values constrained to $(-1,1)^{\dimens}$ to ensure positive-semidefiniteness of $\boldsymbol{\Sigma}_c$.\footnote{Note that $\boldsymbol{S}_c$ is the off-diagonal block of $\boldsymbol{\Sigma}_c$, and it is not required to be neither symmetric nor positive-semidefinite.} \textbf{Case} $\boldsymbol{u}_1 = \boldsymbol{u}_2$: With \glspl{MVN} marginals, this coupling is equivalent to setting $\boldsymbol{z}_1 = \boldsymbol{z}_2$ as discussed above on how to sample from $\mathbb{C}_{1:2}^{\texttt{MVN}}$. Therefore, by simply setting $\boldsymbol{S}_c = \boldsymbol{I}_{\dimens}$, this coupling is recovered as a special case, in which case  $\boldsymbol{M} \boldsymbol{M}^\top = \boldsymbol{I}_{\dimens} - \boldsymbol{S}_{c} \boldsymbol{S}_{c}^\top = \boldsymbol{0}$.  \textbf{Case} $\boldsymbol{u}_1 = \boldsymbol{1} - \boldsymbol{u}_2$: this is a straightforward extension of the previous case. Indeed, setting $\boldsymbol{S}_c$ to an orthogonal matrix generalizes both these cases and includes other new variants. In \citep{henderson2012sharpening}, this technique was briefly mentioned, but not elaborated.

\paragraph{Gradient estimation.}
Throughout this part we use the notation $\omega$ defined in Section~\ref{sec:optimalcoupling}. Recall that the marginals $q_1,q_2$ are assumed fixed. Only the coupling parameter $\thetab_c$ is a free variable in the below discussion. As discussed in the main paper, we need to know the sign of $I$, and this is usually not a big issue in practice from a pilot run. 

Then we have two options for gradient estimation based on the differentiability of $\varphi$. The first one is pathwise gradient estimation and it requires differentiable $\varphi$ w.r.t. to its inputs. It works as follows: obtain a pair of base variables $(\n_1,\n_2)\sim\nu$, whose distribution does not depend on $\thetab_c$. Then these are transformed to a coupled sample by $(\boldsymbol{u}_1,\boldsymbol{u}_2)=(\boldsymbol{R}_1(\n_1;\thetab_c),\boldsymbol{R}_2(\n_2;\thetab_c))$. \footnote{The second-moment assumptions in Proposition~\ref{thm:asym_var} of the main paper make $\omega$ integrable by Cauchy--Schwarz.} We can then write the cross-moment objective as an expectation under the base distribution $\nu$, 
\begin{equation}\label{eq:objective_reparam}
\mathcal C(\thetab_c)
=
\mathbb{E}_\nu\!\left[
\omega\!\left(\boldsymbol{R}_1(\n_1;\thetab_c),
\boldsymbol{R}_2(\n_2;\thetab_c)\right)
\right].
\end{equation}
 Then, pathwise gradient estimator is
\begin{equation}\label{eq:gradient_mc_estimate}
\widehat{\nabla}_{\thetab_c}^{M,\mathrm{Path}}\mathcal C(\thetab_c)
=
\frac1M\sum_{m=1}^M
\nabla_{\thetab_c}\!\left[
\omega\!\left(\boldsymbol{R}_1(\n_1^{(m)};\thetab_c),
\boldsymbol{R}_2(\n_2^{(m)};\thetab_c)\right)
\right],
\end{equation}
 It is unbiased under mild regularity conditions on differentiating under the integral sign For $\varphi\geq0$.  

Let us now discuss the other gradient estimator, i.e., the score function estimator. It reqyures a density for the coupling $c_{1:2}(\boldsymbol{u}_1,\boldsymbol{u}_2;\thetab_c)$. The score-function estimator is given by
$$
\widehat{\nabla}_{\thetab_c}^{M,\mathrm{Score}}\mathcal C(\thetab_c)
=
\frac1M\sum_{m=1}^M
\omega(\boldsymbol{u}_1^{(m)},\boldsymbol{u}_2^{(m)})
\nabla_{\thetab_c}
\log c_{1:2}(\boldsymbol{u}_1^{(m)},\boldsymbol{u}_2^{(m)};\thetab_c),
$$
where the samples are i.i.d.\ from $c_{1:2}(\,\cdot\,;\thetab_c)$. The advantage is that we do not need $\varphi$ to be differentiable w.r.t. to its inputs. 

\section{Variance derivations for GenSNIS}\label{app:deriv}
Here we shall prove the variance formulas in Propositions~\ref{prop:snis-asym-var} and~\ref{thm:asym_var} of the main paper. They both follow from the same expression for the estimation error which also explains the cancellation result in Section~\ref{sec:theory}.

\paragraph{Deriving the asymptotic variance.}
Recall $I$, $H$ and the optimal UIS proposals from Proposition~\ref{prop:snis-asym-var}. Assume now $q_i^{\bigstar}\ll q_i$ for $i=1,2$. For a sample $(\boldsymbol{x}_1,\boldsymbol{x}_2)\sim\jointprop$, write
$$
A=\frac{\varphi(\boldsymbol{x}_1)\pi(\boldsymbol{x}_1)}{q_1(\boldsymbol{x}_1)},
\qquad
V=\frac{\pi(\boldsymbol{x}_2)}{q_2(\boldsymbol{x}_2)},
\qquad r=A-IV.
$$
Thus $r$ is the residual introduced in Section~\ref{sec:theory}. The support assumptions give $\mathbb{E}A=I$, $\mathbb{E}|A|=H$ and $\mathbb{E}V=1$, so $r$ is integrable with mean zero. For $N$ i.i.d.\ pairs, let $\overline r_N=N^{-1}\sum_{n=1}^N r_n$ and $\overline V_N=N^{-1}\sum_{n=1}^N V_n=\hatz/Z_\pi$. Whenever $\hatz>0$, the error is exactly
\begin{equation}\label{eq:app-residual-identity}
\GenSNISest-I
=\frac{\hatj-I\hatz}{\hatz}
=\frac{\overline r_N}{\overline V_N}.
\end{equation}
If $\mathbb{E}r^2<\infty$, the central limit theorem applies to $\sqrt N\,\overline r_N$. Since $\overline V_N\to1$ almost surely by the strong law of large numbers, Slutsky's theorem gives
\begin{equation}\label{eq:app-residual-clt}
\sqrt N(\GenSNISest-I)
\xrightarrow{d}\mathcal N(0,\mathbb{E}_{\jointprop}r^2).
\end{equation}
The zero-denominator convention does not affect this limit: its probability is $\mathbb P(V=0)^N\to0$. This argument only needs a finite second moment of the residual. Section~\ref{app:finite-variance-coupling} gives an example where neither separate weight has one.

\paragraph{Proof of Propositions~\ref{prop:snis-asym-var} and~\ref{thm:asym_var}.}
Under their separate second-moment assumptions, we can expand $\mathbb{E}r^2$ into three finite terms. The identity $A=H\operatorname{sign}\{\varphi(\boldsymbol{x}_1)\}q_1^{\bigstar}(\boldsymbol{x}_1)/q_1(\boldsymbol{x}_1)$ gives
$$
\mathbb{E}[A^2]=H^2\{1+\chi^2(q_1^{\bigstar}\|q_1)\},
\qquad
\mathbb{E}[V^2]=1+\chi^2(q_2^{\bigstar}\|q_2).
$$
Substitution into $\mathbb{E}(A-IV)^2$ therefore yields
\begin{equation}\label{eq:app-general-variance}
\begin{aligned}
\mathbb{V}_{\jointprop}^{\infty}[\GenSNISest]
={}&H^2\{1+\chi^2(q_1^{\bigstar}\|q_1)\}
+I^2\{1+\chi^2(q_2^{\bigstar}\|q_2)\}\\
&-2IH\,\mathbb{E}_{\jointprop}\!\left[
\operatorname{sign}\{\varphi(\boldsymbol{x}_1)\}
\frac{q_1^{\bigstar}(\boldsymbol{x}_1)q_2^{\bigstar}(\boldsymbol{x}_2)}
{q_1(\boldsymbol{x}_1)q_2(\boldsymbol{x}_2)}
\right].
\end{aligned}
\end{equation}
This proves Proposition~\ref{thm:asym_var}. Taking $\boldsymbol{x}_1=\boldsymbol{x}_2\sim q$ proves Proposition~\ref{prop:snis-asym-var}. With fixed proposal marginals, only the cross moment can change, and its coefficient $-2IH$ determines the optimization direction. The relation between this asymptotic variance and the estimator's mean squared error requires further control of small denominator estimates, supplied in Section~\ref{app:negative-moment-denominator}.

\paragraph{The remaining numerator variance for signed test functions.}\label{app:signed-test-functions}
The optimal proposal $q_1^{\bigstar}$ removes variation in the magnitude of the numerator weight, but its sign may still vary. Writing $\varphi_\pm=\max(\pm\varphi,0)$ and $I_\pm=\mathbb{E}_\pi\varphi_\pm$, we have $I=I_+-I_-$ and $H=I_++I_-$. Hence
$$
\operatorname{Var}_{q_1}(A)
=H^2\chi^2(q_1^{\bigstar}\|q_1)+H^2-I^2,
\qquad H^2-I^2=4I_+I_-.
$$
At $q_1=q_1^{\bigstar}$, this variance remains positive if both $I_+$ and $I_-$ are positive. The decomposition into positive and negative parts also appears in \citet[Section~4.3]{li2013twostage}. In \citet[Section~3.2]{rainforth2020target}, they proposed to use positivisation \citep{owen2000safe} to split the numerator integral in a difference of two integrals, each admitting a zero-variance proposal once using a UIS estimator. We did not pursue this extension, as then it would also be natural to couple these two estimators inside the numerator itself, complicating the framework further.

\paragraph{A lower bound under separate second moments.}\label{app:proof}
Coupling helps when a large fluctuation in $A$ is accompanied by a similar fluctuation in $IV$. But the sizes of these fluctuations are constrained by the chosen marginals. If their standard deviations differ, some variance must remain, however we pair the samples.
\begin{proposition}\label{prop:variance-lower-bound}
If $q_i^{\bigstar}\ll q_i$ and $q_i^{\bigstar}/q_i\in L^2(q_i)$ for $i=1,2$, every coupling with these marginals satisfies
\begin{equation}\label{eq_max_cov}
\mathbb{V}_{\jointprop}^{\infty}[\GenSNISest]
\geq\left\{
\sqrt{H^2\{1+\chi^2(q_1^{\bigstar}\|q_1)\}-I^2}
-|I|\sqrt{\chi^2(q_2^{\bigstar}\|q_2)}
\right\}^{2}.
\end{equation}
\end{proposition}
\begin{proof}
Cauchy--Schwarz gives $\operatorname{Cov}(A,IV)\leq\sqrt{\operatorname{Var}(A)\operatorname{Var}(IV)}$. Therefore,
$$
\mathbb{E}r^2=\operatorname{Var}(A-IV)
\geq\{\sqrt{\operatorname{Var}(A)}-\sqrt{\operatorname{Var}(IV)}\}^{2}.
$$
Using $\operatorname{Var}(A)=H^2\{1+\chi^2(q_1^{\bigstar}\|q_1)\}-I^2$ and $\operatorname{Var}(IV)=I^2\chi^2(q_2^{\bigstar}\|q_2)$ proves the bound. When both variances are positive, equality requires $A-I$ and $I(V-1)$ to be positively proportional almost surely, which need not be possible for the given marginal distributions. If either variance is zero, equality holds for every coupling.
\end{proof}

\subsection{A non-asymptotic error bound}\label{app:proof_ub}
We prove the bound in Section~\ref{sec:theory} by working with the difference $\hatj-I\hatz$. A coupling can make the error in $\hatj$ nearly cancel with $I$ times the error in $\hatz$, even when each is large.

We retain the independent pairs from $\jointprop$, the support assumptions $q_i^{\bigstar}\ll q_i$ for $i=1,2$, and $\varphi\in L^1(\pi)$. Recall that the residual satisfies
$$
Z_\pi r(\boldsymbol{x}_1,\boldsymbol{x}_2)
=\frac{\varphi(\boldsymbol{x}_1)\widetilde\pi(\boldsymbol{x}_1)}{q_1(\boldsymbol{x}_1)}
-I\frac{\widetilde\pi(\boldsymbol{x}_2)}{q_2(\boldsymbol{x}_2)}.
$$
The support assumptions give $\mathbb{E}\hatj=J=IZ_\pi$ and $\mathbb{E}\hatz=Z_\pi$. Thus $\mathbb{E}_{\jointprop}r=0$, and the triangle inequality gives $\mathbb{E}_{\jointprop}|r|\leq H+|I|<\infty$. Write $r_n=r(\boldsymbol{x}_1^{(n)},\boldsymbol{x}_2^{(n)})$ and $\overline r_N=N^{-1}\sum_{n=1}^N r_n$. On $\hatz>0$, the error identity \eqref{eq:app-residual-identity} becomes
\begin{equation}\label{eq:snisesterror}
\GenSNISest-I
=\frac{\hatj-I\hatz}{\hatz}
=\frac{Z_\pi\overline r_N}{\hatz}.
\end{equation}
We consider $\mathbb{E}_{\jointprop}|r|>0$, as in the theorem. If it is zero, the estimator is exact whenever its denominator is positive.

\paragraph{Keeping the cancellation in the error bound.}
It is useful to pause at the choice of what to bound. In the proof of their SNIS result, \citet[Theorem~1.2]{chatterjee2018sample} separate the errors in the two integrals. In our notation, their triangle inequality is
$$
|\hatj-I\hatz|
=| (\hatj-J)-I(\hatz-Z_\pi) |
\leq|\hatj-J|+|I|\,|\hatz-Z_\pi|.
$$
They bound these errors separately so the resulting bound loses the cancellation. We view this as some information loss, as we know that dependence between the errors does affect the overall error. We will instead bound the average residual in \eqref{eq:snisesterror} directly.

We also need to deal with small denominators $\hatz$, which intuitively in \Cref{eq:snisesterror} can amplify errors to become arbitrarily large if not controlled. 
We shall follow the approach of \citet[Appendix~A.2.2, proof of Theorem~2.2]{deligiannidis2024importance} (who derive different results for SNIS). We first set aside the event $\hatz<Z_\pi/2$. On its complement,
$$
|\GenSNISest-I|
\leq\frac{2}{Z_\pi}|\hatj-I\hatz|
=2|\overline r_N|.
$$
An error larger than $\epsilon>0$ therefore requires either a denominator below $Z_\pi/2$ or an average residual with absolute value larger than $\epsilon/2$. The union bound, followed by Markov's inequality, gives
\begin{equation}\label{eq:residual-denominator-split}
\begin{aligned}
\mathbb P(|\GenSNISest-I|>\epsilon)
&\leq\mathbb P(\hatz<Z_\pi/2)
+\mathbb P(|\overline r_N|>\epsilon/2)\\
&\leq\mathbb P(\hatz<Z_\pi/2)
+\frac{2}{\epsilon}\mathbb{E}|\overline r_N|.
\end{aligned}
\end{equation}
Next, we deal with the second term, i.e., the mean absolute residual, and we will come back to the probability of a small denominator later. 

\paragraph{Upper bounding the mean absolute residual.}
As we said, the residual itself may have an infinite second moment, and this is in fact often the interesting regime. To handle this, we adapt the truncation argument in the proof of Theorem~1.1 of \citet{chatterjee2018sample}. Choose $K>0$ and split each residual according to whether $|r_n|\leq K$. The bounded part $h_n:=r_n\mathbf1_{\{|r_n|\leq K\}}$ has a finite variance, so it is easy to control. Then we will handle the tail values in a separate term.

 Since $\mathbb{E}r=0$, we have $\mathbb{E}h_1=-\mathbb{E}[r\mathbf1_{\{|r|>K\}}]$, and hence we can rewrite 
$$
\overline r_N
=\frac1N\sum_{n=1}^N(h_n-\mathbb{E}h_1)
+\frac1N\sum_{n=1}^N r_n\mathbf1_{\{|r_n|>K\}}
+\mathbb{E}h_1.
$$
For the first term, Cauchy--Schwarz and independence across samples give
$$
\mathbb{E}\left|\frac1N\sum_{n=1}^N(h_n-\mathbb{E}h_1)\right|
\leq\sqrt{\frac{\operatorname{Var}(h_1)}{N}}
\leq\sqrt{\frac{K\mathbb{E}|r|}{N}},
$$
where the last inequality uses $h_1^2\leq K|r_1|$. The expected absolute value of the sampled tail average is at most $\mathbb{E}[|r|\mathbf1_{\{|r|>K\}}]$, by the triangle inequality. The correction $|\mathbb{E}h_1|$ has the same bound. Adding these contributions yields
\begin{align}\label{eq:main_bound_expectedresidual}
\mathbb{E}|\overline r_N|
\leq\underbrace{\sqrt{\frac{K\mathbb{E}|r|}{N}}}_{\text{bound for the truncated average}}
+\underbrace{2\mathbb{E}[|r|\mathbf1_{\{|r|>K\}}]}_{\text{contribution from the tail}}.
\end{align}
Let us notice the role of the cutoff $K$ here: a larger $K$ makes the second term smaller, while it increases the first term. The first term though can also be controlled with larger $N$, i.e., more samples. 

\paragraph{Choosing the cutoff $K$.}
Recall the distribution $\mathbb Q_r$ from the main text, defined as having importance weight (in extended space) $d\mathbb Q_r/d\jointprop=|r|/\mathbb{E}_{\jointprop}|r|$: it gives more probability to samples from the joint with larger absolute residuals. Now notice that the tail term can be written as 
$$
\mathbb{E}[|r|\mathbf1_{\{|r|>K\}}]
=(\mathbb{E}|r|)\,\mathbb Q_r(|r|>K).
$$
We will select the cutoff $K = a \mathbb{E}_{\jointprop}|r|$: this is so that $\mathbb Q_r(|r|>K)$ measures whether the absolute residual exceeds \(a\) times its mean, analogously to how \citep{chatterjee2018sample} uses the standard (log) importance weights. We will choose the value of \(a\) below. Plugging this choice into the bound in \cref{eq:main_bound_expectedresidual}, we obtain

\begin{equation}\label{eq:residual-truncation-bound}
\mathbb{E}|\overline r_N|
\leq(\mathbb{E}|r|)\left\{\sqrt{\frac aN}
+2\mathbb Q_r\left(\frac{|r|}{\mathbb{E}|r|}>a\right)\right\}.
\end{equation}
Now, to obtain a sufficient sample size, we can choose $a = e^{L_r+t/2}$, with $t\geq0$, and (assuming finiteness of $L_r=\mathrm{KL}(\mathbb Q_r\|\jointprop)$, which we comment on soon) when $N\geq e^{L_r+t}$, we straightforwardly see from \Cref{eq:residual-truncation-bound} that
$$
\sqrt{\frac aN}\leq e^{-t/4},
\qquad
\left\{\frac{|r|}{\mathbb{E}|r|}>a\right\}
=\left\{\log\frac{|r|}{\mathbb{E}|r|}>L_r+\frac t2\right\}.
$$
Subsequently then substituting into \eqref{eq:residual-truncation-bound} and \eqref{eq:residual-denominator-split} finally gives
$$
\begin{aligned}
\mathbb P(|\GenSNISest-I|>\epsilon)
\leq{}&\mathbb P(\hatz<Z_\pi/2)\\
&+\frac{2\mathbb{E}|r|}{\epsilon}
\left\{e^{-t/4}
+2\mathbb Q_r\left(\log\frac{|r|}{\mathbb{E}|r|}>L_r+\frac t2\right)\right\},
\end{aligned}
$$
which proves the theorem.

\paragraph{Bounding the probability of a small denominator.}
It now remains to justify the exponential bound on this probability that we have briefly mentioned in the main text. The denominator is an average of independent nonnegative weights with mean $Z_\pi>0$. For any $\lambda>0$, applying Markov's inequality to $e^{-\lambda N\hatz}$ gives
$$
\begin{aligned}
\mathbb P(\hatz<Z_\pi/2)
&\leq e^{\lambda NZ_\pi/2}\mathbb{E}[e^{-\lambda N\hatz}]\\
&=\left\{e^{\lambda Z_\pi/2}\mathbb{E}_{q_2}
\exp\left[-\lambda\frac{\widetilde\pi(\boldsymbol{x}_2)}{q_2(\boldsymbol{x}_2)}\right]\right\}^N,
\end{aligned}
$$
Call the expression in braces $f(\lambda)$. Its value at zero is one, and
$$
f'(0+)=\frac{Z_\pi}{2}
-\mathbb{E}_{q_2}\left[\frac{\widetilde\pi(\boldsymbol{x}_2)}{q_2(\boldsymbol{x}_2)}\right]
=-\frac{Z_\pi}{2}<0.
$$
The derivative follows from $\mathbb{E}_{q_2}[\widetilde\pi/q_2]=Z_\pi$ and $0\leq(1-e^{-\lambda w})/\lambda\leq w$ for $w\geq0$. Thus some $\lambda>0$ gives $\kappa:=f(\lambda)<1$, and $\mathbb P(\hatz<Z_\pi/2)\leq\kappa^N$.

We can now give the sufficient sample size stated in the main paper. Fix $\epsilon>0$ and $\delta\in(0,1)$, and write $m_r=\mathbb{E}_{\jointprop}|r|$. Choose $t\geq0$ so that
$$
e^{-t/4}+2\mathbb Q_r\left(\log\frac{|r|}{m_r}>L_r+\frac t2\right)
\leq\frac{\delta\epsilon}{4m_r}.
$$
Such a choice exists because the upper-tail probability tends to zero as $t$ increases. For $N\geq e^{L_r+t}$, the residual term in the bound is at most $\delta/2$. The denominator term is also at most $\delta/2$ when $N\geq\log(2/\delta)/(-\log\kappa)$. Hence
$$
N\geq\max\left\{e^{L_r+t},\frac{\log(2/\delta)}{-\log\kappa}\right\}
\quad\Longrightarrow\quad
\mathbb P(|\GenSNISest-I|>\epsilon)\leq\delta.
$$
\subsubsection{A lower bound and a necessary sample size}\label{app:residual-lower-bound}
The sufficient sample size above ensures that the probability of an error larger than $\epsilon$ is at most $\delta$. We now seek sample sizes for which this probability must exceed $\delta$. If $N$ is small, we may miss large values of $|r|$. We need to establish when missing those values makes $\GenSNISest$ inaccurate. To do so, we bound the probability that $\GenSNISest$ is still at least $I-\epsilon$ when all sampled values of $|r|$ lie below a cutoff.

Recall the support and integrability assumptions of Section~\ref{app:proof_ub}, and suppose $0<\mathbb{E}_{\jointprop}|r|<\infty$. We additionally assume $\widetilde\pi(\boldsymbol{x}_2)/q_2(\boldsymbol{x}_2)>0$ almost surely under $q_2$. This is stronger than the condition used for the upper bound and ensures that $\hatz>0$ almost surely. We use this positivity below to rewrite $\GenSNISest\geq I-\epsilon$ as an equivalent inequality involving $\overline r_N$ and $\hatz$. We write $\mathbb P_{\jointprop^{\otimes N}}$ for probabilities over the $N$ i.i.d.\ samples from $\jointprop$. Expectations $\mathbb{E}_{\jointprop}$ are over one sample $(\boldsymbol{x}_1,\boldsymbol{x}_2)\sim\jointprop$, and $r$ inside such an expectation means $r(\boldsymbol{x}_1,\boldsymbol{x}_2)$. For a real number $u$, write $u_+=\max(u,0)$.

\textbf{Lower bound.} For every $a>0$, $\epsilon>0$ and positive integer $N$,
\begin{equation}\label{eq:residual-error-lower-bound}
\begin{aligned}
\mathbb P_{\jointprop^{\otimes N}}\bigl(|\GenSNISest-I|>\epsilon\bigr)
\geq{}&1-\frac Na\\
&-\frac{2}{\mathbb{E}_{\jointprop}|r|}
\mathbb{E}_{\jointprop}\!\left[
\left(r+\epsilon\frac{\widetilde\pi(\boldsymbol{x}_2)}{Z_\pi q_2(\boldsymbol{x}_2)}
+\frac{\mathbb{E}_{\jointprop}|r|}{2}\right)_+
\mathbf1_{\{|r|\leq a\mathbb{E}_{\jointprop}|r|\}}
\right].
\end{aligned}
\end{equation}
The term $N/a$ bounds the probability that at least one sampled value of $|r|$ exceeds the cutoff. The second subtracted term bounds the probability that $\GenSNISest\geq I-\epsilon$ while every sampled value of $|r|$ stays below the cutoff. When both terms are small, underestimation by more than $\epsilon$ has high probability.

\begin{proof}
First choose $a>0$ and use the cutoff $a\mathbb{E}_{\jointprop}|r|$, as in the upper-bound argument. Recall $r_n=r(\boldsymbol{x}_1^{(n)},\boldsymbol{x}_2^{(n)})$. The union bound and Markov's inequality give
\begin{equation}\label{eq:residual-lower-cutoff}
\begin{aligned}
\mathbb P_{\jointprop^{\otimes N}}\!\left(\max_{1\leq n\leq N}|r_n|>a\mathbb{E}_{\jointprop}|r|\right)
&\leq N\jointprop\!\left(|r|>a\mathbb{E}_{\jointprop}|r|\right)\\
&\leq\frac Na.
\end{aligned}
\end{equation}
We next consider samples for which every $|r_n|$ is at most $a\mathbb{E}_{\jointprop}|r|$. To relate those samples to the accuracy of $\GenSNISest$, use the error identity \eqref{eq:snisesterror}. Since $\hatz>0$,
\begin{equation}\label{eq:residual-lower-error-event}
\GenSNISest\geq I-\epsilon
\quad\Longleftrightarrow\quad
\overline r_N+\epsilon\frac{\hatz}{Z_\pi}\geq0.
\end{equation}
The expression on the right is an average of $r_n+\epsilon\widetilde\pi(\boldsymbol{x}_2^{(n)})/\{Z_\pi q_2(\boldsymbol{x}_2^{(n)})\}$. We can bound the probability that this average is nonnegative by adding a positive constant and applying Markov's inequality to positive parts. A natural constant is $\mathbb{E}_{\jointprop}|r|/2$: because $\mathbb{E}_{\jointprop}r=0$, we have
$$
\mathbb{E}_{\jointprop}(r)_+
=\mathbb{E}_{\jointprop}(-r)_+
=\frac{\mathbb{E}_{\jointprop}|r|}{2}.
$$
On the event $\GenSNISest\geq I-\epsilon$, adding this constant to each term in the average in \eqref{eq:residual-lower-error-event} makes the average at least $\mathbb{E}_{\jointprop}|r|/2$. Replacing each term by its positive part can only increase the average. Therefore, if $\GenSNISest\geq I-\epsilon$ and every $|r_n|\leq a\mathbb{E}_{\jointprop}|r|$, then
$$
\frac1N\sum_{n=1}^N
\left(r_n+\epsilon\frac{\widetilde\pi(\boldsymbol{x}_2^{(n)})}{Z_\pi q_2(\boldsymbol{x}_2^{(n)})}
+\frac{\mathbb{E}_{\jointprop}|r|}{2}\right)_+
\mathbf1_{\{|r_n|\leq a\mathbb{E}_{\jointprop}|r|\}}
\geq\frac{\mathbb{E}_{\jointprop}|r|}{2}.
$$
The indicators equal one on the event being considered. Markov's inequality applied to the nonnegative average now gives
\begin{equation}\label{eq:residual-lower-retained}
\begin{aligned}
&\mathbb P_{\jointprop^{\otimes N}}\!\left(
\GenSNISest\geq I-\epsilon,\quad
\max_{1\leq n\leq N}|r_n|\leq a\mathbb{E}_{\jointprop}|r|
\right)\\
&\qquad\leq\frac{2}{\mathbb{E}_{\jointprop}|r|}
\mathbb{E}_{\jointprop}\!\left[
\left(r+\epsilon\frac{\widetilde\pi(\boldsymbol{x}_2)}{Z_\pi q_2(\boldsymbol{x}_2)}
+\frac{\mathbb{E}_{\jointprop}|r|}{2}\right)_+
\mathbf1_{\{|r|\leq a\mathbb{E}_{\jointprop}|r|\}}
\right].
\end{aligned}
\end{equation}
All $N$ terms have the same expectation, which is why the right-hand side is an expectation over one sample $(\boldsymbol{x}_1,\boldsymbol{x}_2)\sim\jointprop$. The expectation is finite: its integrand is at most $|r|+\epsilon\widetilde\pi(\boldsymbol{x}_2)/\{Z_\pi q_2(\boldsymbol{x}_2)\}+\mathbb{E}_{\jointprop}|r|/2$, and $\mathbb{E}_{q_2}[\widetilde\pi/(Z_\pi q_2)]=1$.

Finally, split the event $\{\GenSNISest\geq I-\epsilon\}$ according to whether any $|r_n|$ exceeds the cutoff. Equations~\eqref{eq:residual-lower-cutoff} and~\eqref{eq:residual-lower-retained} bound the probabilities of the two parts. Taking complements gives the lower bound for $\mathbb P_{\jointprop^{\otimes N}}(\GenSNISest<I-\epsilon)$. Since underestimation by more than $\epsilon$ implies $|\GenSNISest-I|>\epsilon$, \eqref{eq:residual-error-lower-bound} follows.
\end{proof}
Replacing $r$ by $-r$ in the expectation in \eqref{eq:residual-error-lower-bound} gives the corresponding bound through overestimation by more than $\epsilon$.

\paragraph{Choosing the cutoff and obtaining a necessary sample size.}
Fix $\epsilon>0$ and $\delta\in(0,1)$. We want to identify sample sizes that cannot satisfy $\mathbb P_{\jointprop^{\otimes N}}(|\GenSNISest-I|>\epsilon)\leq\delta$. For a fixed $a$, the expectation in \eqref{eq:residual-error-lower-bound} is independent of $N$. The only dependence on $N$ of the whole expression is via $-N/a$: a lower bound above $\delta$ at a specific sample size also rules out every other \emph{smaller} sample size.

To express such sample sizes using the same divergence as in the upper bound, assume $L_r=\operatorname{KL}(\mathbb Q_r\|\jointprop)<\infty$. We follow the cutoff choice in the proof of Theorem~1.2 of \citet{chatterjee2018sample}, now applied to $|r|/\mathbb{E}_{\jointprop}|r|$, and choose
$$
a=e^{L_r-t/2},\qquad t\geq0.
$$
For every positive integer $N\leq e^{L_r-t}$,
$$
\frac Na\leq\frac{e^{L_r-t}}{e^{L_r-t/2}}=e^{-t/2}.
$$
The remaining requirement is that the expectation in \eqref{eq:residual-error-lower-bound} be small enough. Specifically, suppose that the chosen $t$ satisfies
\begin{equation}\label{eq:residual-necessary-condition}
\frac{2}{\mathbb{E}_{\jointprop}|r|}
\mathbb{E}_{\jointprop}\!\left[
\left(r+\epsilon\frac{\widetilde\pi(\boldsymbol{x}_2)}{Z_\pi q_2(\boldsymbol{x}_2)}
+\frac{\mathbb{E}_{\jointprop}|r|}{2}\right)_+
\mathbf1_{\{|r|\leq a\mathbb{E}_{\jointprop}|r|\}}
\right]
<1-\delta-e^{-t/2}.
\end{equation}
Substituting $N/a\leq e^{-t/2}$ and \eqref{eq:residual-necessary-condition} into \eqref{eq:residual-error-lower-bound} gives
$$
\mathbb P_{\jointprop^{\otimes N}}\bigl(|\GenSNISest-I|>\epsilon\bigr)>\delta
\qquad\text{for every positive integer }N\leq e^{L_r-t}.
$$
Thus, assuming \eqref{eq:residual-necessary-condition} holds,
\begin{equation}\label{eq:residual-necessary-size}
\mathbb P_{\jointprop^{\otimes N}}\bigl(|\GenSNISest-I|>\epsilon\bigr)\leq\delta
\quad\Longrightarrow\quad N>e^{L_r-t}.
\end{equation}
The condition~\eqref{eq:residual-necessary-condition} is an additional requirement on the given $\varphi$, $\jointprop$. Together with $N\leq e^{L_r-t}$, that condition makes both subtracted terms in \eqref{eq:residual-error-lower-bound} small enough so that the error probability is above $\delta$. Instead our lower bound in \eqref{eq:residual-error-lower-bound} does not require this condition, it uses only first moments. 

\subsection{An optimal coupling for factorized proposals}\label{app:crn}
In this Section, we present precise conditions under which the CRN coupling from Section~\ref{sec:paramcoupling} is optimal. With factorized proposals, the one-dimensional rearrangement inequality can be applied successively to matching coordinates. This is provided their samples are independent. In this way we obtain an optimality result, for a specific class of couplings. We also present a quick example to demonstrate the relevance of this result.

\begin{proposition}[Asymptotically optimal couplings: a special case]\label{prop:asymp_opt_coupling_special_case}\label{prop:crn-coordinate-class}
    Let $d\in\mathbb{N}$. Assume the two marginal proposals are factorized, i.e., 
    $$
    q_{1}(\boldsymbol{x}_1)=\prod_{j=1}^d q_{1}^{(j)}(x_{1,j}),
    \qquad
    q_{2}(\boldsymbol{x}_2)=\prod_{j=1}^d q_{2}^{(j)}(x_{2,j}).
    $$
    Let $\boldsymbol{T}_1,\boldsymbol{T}_2:[0,1]^d\to\mathbb{R}^d$ be the corresponding componentwise inverse-CDF maps,
    so that $\boldsymbol{T}_i(\boldsymbol{u}_i)=(T_i^{(1)}(u_{i,1}),\dots,T_i^{(d)}(u_{i,d}))$ with each $T_i^{(j)}$ nondecreasing, and
    $\boldsymbol{T}_i\sharp \mathrm{Unif}([0,1]^d)=\mathbb{Q}_i$ for $i\in\{1,2\}$.
    
    Then recall:
    $$
    w_1(\boldsymbol{x}_1)=\frac{\varphi(\boldsymbol{x}_1)\,\pi(\boldsymbol{x}_1)}{q_1(\boldsymbol{x}_1)},
    \qquad
    w_2(\boldsymbol{x}_2)=I\frac{\pi(\boldsymbol{x}_2)}{q_2(\boldsymbol{x}_2)},
    $$
    and express them in terms of the inputs $(\boldsymbol u_1, \boldsymbol u_2) ~ \text{Uniform}$, as 
    $$
    g^{\star}(\boldsymbol{u}_1)=w_1(\boldsymbol{T}_1(\boldsymbol{u}_1)),
    \qquad
    h^{\star}(\boldsymbol{u}_2)=w_2(\boldsymbol{T}_2(\boldsymbol{u}_2)),
    \qquad \boldsymbol{u}_1,\boldsymbol{u}_2\in[0,1]^d.
    $$
    Now assume $g^{\star},h^{\star}\in L^2([0,1]^d)$. Then for each coordinate, assume that these two functions are monotone in one common direction, uniformly over the other coordinates. This condition holds, for instance, when $w_1$ and $w_2$ have the same coordinatewise monotonicity.
    
    Let $\mathcal{O}_d$ be the class of couplings $\mathbb{C}_{1:2}$ of $(\boldsymbol{U}_1,\boldsymbol{U}_2)\in[0,1]^d\times[0,1]^d$ such that
    (i) $\boldsymbol{U}_1\sim \mathrm{Unif}([0,1]^d)$ and $\boldsymbol{U}_2\sim \mathrm{Unif}([0,1]^d)$, and
    (ii) dependence is permitted only between like components, i.e.\ the $d$ bivariate pairs
    $\{(U_{1,j},U_{2,j})\}_{j=1}^d$ are mutually independent.\footnote{This is porecisely the admissible class of couplings
    considered in Proposition~1 of the Appendix of \citep{rubinstein1985variance}.}
    
    Then we can show that the \gls{CRN} coupling $\boldsymbol{U}_1=\boldsymbol{U}_2=\boldsymbol{U}$ maximizes their product expectation within $\mathcal O_d$:
    $$
    \max_{\mathbb{C}_{1:2}\in\mathcal{O}_d}
    \mathbb{E}_{\mathbb{C}_{1:2}}[g^{\star}(\boldsymbol{U}_1)h^{\star}(\boldsymbol{U}_2)]
    =\mathbb{E}[g^{\star}(\boldsymbol{U})h^{\star}(\boldsymbol{U})],
    \qquad \boldsymbol{U}\sim\mathrm{Unif}([0,1]^d).
    $$
    Equivalently, among such couplings the CRN choice minimizes the asymptotic variance of GenSNIS.
    \end{proposition}
    
    \begin{proof}
    Fix a coordinate $j$ and a coupling in $\mathcal O_d$. By definition, $(U_{1,j},U_{2,j})$ is independent of all the other coordinate pairs. We can therefore condition on those pairs, and compare
    the coupling in the remaining coordinate. Under this conditioning, $g^{\star}$ and
    $h^{\star}$ are functions of $U_{1,j}$ and $U_{2,j}$, respectively,
    that are monotone in the same direction. The one-dimensional
    rearrangement inequality then shows that their conditional product
    expectation cannot decrease when this coordinate pair is replaced
    by $(U_j,U_j)$, with $U_j$ being uniform. Taking expectations preserves this comparison.
    Because the replacement preserves membership in $\mathcal O_d$, we can repeat it for $j=1,\ldots,d$. The resulting coupling is $\boldsymbol{U}_1=\boldsymbol{U}_2=\boldsymbol{U}$, which proves the claim. Cauchy--Schwarz and the $L^2$ assumptions on $g^{\star}$ and $h^{\star}$ ensure that all these cross moments are finite.
    \end{proof}

    Let us present a short example that shows that allowing dependence across different coordinates can improve on CRN. For $d=2$,
    consider the normalized, coordinatewise nondecreasing weight
    functions $g^{\star}(\boldsymbol{u})=2u_1$ and $h^{\star}(\boldsymbol{v})=2v_2$.
    Under CRN, the two weights depend on independent coordinates, so their cross moment is
    $4\mathbb{E}[U_1U_2]=1$. If instead $\boldsymbol{U}=(U_1,U_2)$ has independent uniform
    coordinates and $\boldsymbol{V}=(U_2,U_1)$, both vectors remain uniform and the two weights depend on the same coordinate. Their
    cross moment is then $4\mathbb{E}[U_1^2]=4/3$. This coupling is outside the family $\mathcal O_2$: its coordinate pairs are not independent.

\subsection{Finite variance through coupling: an example}\label{app:finite-variance-coupling}
The observation in Section~\ref{sec:theory} asserts that coupling can restore finite asymptotic variance when independent numerator and denominator estimation has infinite variance. We here give an concrete (abstract) example demonstrating this. We will specifically check that the estimator has finite variance for every $N$, which is stronger than finite asymptotic variance.

\begin{example}\label{ex:coupling-restores-finite-variance}
Let $q$ be the joint density of independent variables $T$ and $\varepsilon$, where $\mathbb P(T>t)=t^{-\kappa}$ for $t\geq1$, $1<\kappa<2$, and $\varepsilon\sim\mathrm{Unif}[-1/2,1/2]$. Write $m_T=\mathbb{E}T=\kappa/(\kappa-1)$, and choose
$$
q_1=q_2=q,\qquad
\pi(t,e)=\frac{t}{m_T}q(t,e),\qquad
\varphi(t,e)=\frac{t+e}{t}.
$$
These define a probability density and a positive integrand with $I=1$. Both proposal support conditions hold. For a sample $(T,\varepsilon)$ from $q$, the normalized numerator and denominator weights are
$$
A=\frac{T+\varepsilon}{m_T},\qquad V=\frac{T}{m_T}.
$$
Each has an infinite second moment because $\mathbb{E}T^2=\infty$ and $\varepsilon$ is bounded. With independent samples for the two weights, a large numerator weight need not be accompanied by a large denominator weight, and $\mathbb{E}(A-IV)^2=\infty$.

To make the heavy-tailed terms cancel, we use the same sample $(T,\varepsilon)$ in both weights. Subtracting them leaves only $r=\varepsilon/m_T$, and hence $\mathbb{E}r^2=1/(12m_T^2)$. Equation~\eqref{eq:app-residual-clt} gives
$$
\sqrt N(\GenSNISest-1)
\xrightarrow{d}\mathcal N\!\left(0,\frac{1}{12m_T^2}\right).
$$
In this example the denominator is also bounded away from zero, since every $T_n\geq1$. The exact error and its second moment therefore satisfy
$$
\GenSNISest-1=\frac{\sum_{n=1}^N\varepsilon_n}{\sum_{n=1}^N T_n},
\qquad
\mathbb{E}[(\GenSNISest-1)^2]
\leq\frac{\mathbb{E}[(\sum_{n=1}^N\varepsilon_n)^2]}{N^2}
=\frac{1}{12N}.
$$
By comparison, with independent numerator and denominator samples, the numerator sum has infinite second moment and is independent of the positive denominator sum. The denominator is bounded below by $N/m_T$ so we know that the ratio has finite mean. However, because of independence, also know that its second moment is infinite (for every $N$). 
\end{example}

\subsection{Non-asymptotic MSE expansion under some moment assumptions}
\label{app:negative-moment-denominator}

In the previous sections we discused some asymptotic results, looking at the limiting distribution of the ratio estimator. We now study the finite sample MSE for generalized SNIS, with an interest in how to handle a potentially problematic small denominator. Indeed, for asymptotics and in particular asymptotic variance, we saw in the main text that because the probability of a zero denominator in SNIS (hence generalized SNIS) goes to zero fast enough as $N$ increases (geometrically), it does not cause problems. Here we will need to deal with a too-small denominator differently namely by assuming a finite negative moment of the denominator's importance weight. Under the moment assumptions below, we use the negative-moment bound of \citet{hoyt2024mean} and the Taylor expansion in \citet{owen2025better} to justify the approximation and to bound its remainder. 

We work under the support assumptions and with the i.i.d.\ sample pairs introduced at the start of Section~\ref{app:deriv}. Using the notation there, the squared error in \eqref{eq:app-residual-identity} is $\overline r_N^2/\overline V_N^2$. Our aim is to show that its expectation equals $\mathbb{E}r^2/N$ up to a remainder of order $N^{-2}$.

\begin{proposition}[MSE expansion without a deterministic denominator bound]
\label{prop:negative-moment-mse}
Assume that $V>0$ almost surely and, for some $\alpha>0$,
$$
\mathbb{E}[|A|^6]<\infty,
\qquad
\mathbb{E}[V^6]<\infty,
\qquad
\mathbb{E}[V^{-\alpha}]<\infty.
$$
Then, as $N\to\infty$,
\begin{equation}
\mathbb{E}\left[
\left(
\widehat{I}_{\mathrm{GenSNIS}}^N-I
\right)^2
\right]
=
\frac{1}{N}\mathbb{E}\left[r^2\right]
+
O(N^{-2}).
\label{eq:negative-moment-mse-expansion}
\end{equation}
\end{proposition}

\begin{proof}
We first bound inverse powers of the denominator average, which will occur when we expand the squared error. For any $p>0$,
the arithmetic--geometric mean inequality gives
$$
\overline{V}_N^{-p}
\leq
\prod_{n=1}^N V_n^{-p/N}.
$$
If $N\geq p/\alpha$, independence and Jensen's inequality give
\begin{equation}
\mathbb{E}\left[\overline{V}_N^{-p}\right]
\leq
\left\{
\mathbb{E}\left[V^{-\alpha}\right]
\right\}^{p/\alpha}.
\label{eq:negative-moment-average}
\end{equation}
 Note that this is also the bound in \citet[Proposition~A.2]{deligiannidis2024importance}. Thus a finite negative moment of $V$ lets us bound the expectation of any fixed inverse power of $\overline V_N$, once $N$ is sufficiently large.

Let
$$
\Delta_N:=\overline{V}_N-1.
$$
The squared error in \eqref{eq:app-residual-identity} contains $\overline V_N^{-2}$. To isolate its leading term, expand $x\mapsto x^{-2}$ around the denominator's mean, $x=1$:
$$
\frac{1}{\overline{V}_N^2}
=
1-2\Delta_N
+
3\Delta_N^2\xi_N^{-4},
$$
where $\xi_N$ lies between $1$ and $\overline{V}_N$. Therefore,
\begin{align}
\mathbb{E}\left[
\left(
\widehat{I}_{\mathrm{GenSNIS}}^N-I
\right)^2
\right]
&=
\mathbb{E}\left[\overline{r}_N^2\right]
-2\mathbb{E}\left[\overline{r}_N^2\Delta_N\right]
\nonumber\\
&\quad
+3\mathbb{E}\left[
\overline{r}_N^2\Delta_N^2\xi_N^{-4}
\right].
\label{eq:negative-moment-taylor}
\end{align}

Since $\mathbb{E}[r]=0$,
$$
\mathbb{E}\left[\overline{r}_N^2\right]
=
\frac{\mathbb{E}[r^2]}{N}.
$$
Next, to evaluate $\mathbb{E}[\overline r_N^2\Delta_N]$, expand the averages into a sum over three sample indices. If an index occurs only once, its factor is independent of the other factors and has mean zero, since $\mathbb{E}r=\mathbb{E}(V-1)=0$. Only the $N$ terms with all three indices equal remain, giving
$$
\mathbb{E}\left[\overline{r}_N^2\Delta_N\right]
=
\frac{\mathbb{E}[r^2(V-1)]}{N^2}.
$$

It remains to control the last term in
Eq.~\eqref{eq:negative-moment-taylor}. Since $\xi_N$ lies between
$1$ and $\overline{V}_N$,
$$
\xi_N^{-4}
\leq
1+\overline{V}_N^{-4}.
$$
For a centered average of i.i.d.\ variables with a finite sixth moment, expanding the sixth power gives an $O(N^{-3})$ bound. Applying this to $r_n$ and $V_n-1$ yields
$$
\mathbb{E}\left[|\overline{r}_N|^6\right]
=
O(N^{-3}),
\qquad
\mathbb{E}\left[|\Delta_N|^6\right]
=
O(N^{-3}).
$$
Moreover, by H\"older's inequality,
$$
\begin{aligned}
\mathbb{E}\left[
\overline{r}_N^2\Delta_N^2\overline{V}_N^{-4}
\right]
&\leq
\mathbb{E}\left[|\overline{r}_N|^6\right]^{1/3}
\mathbb{E}\left[|\Delta_N|^6\right]^{1/3}
\mathbb{E}\left[\overline{V}_N^{-12}\right]^{1/3}.
\end{aligned}
$$
The first two factors are each $O(N^{-1})$, and Eq.~\eqref{eq:negative-moment-average} bounds the last factor for $N\geq12/\alpha$. We can apply the same bound (with the last factor replaced by one) to control $\mathbb{E}[\overline r_N^2\Delta_N^2]$. Therefore, the full Taylor remainder is $O(N^{-2})$, which proves
Eq.~\eqref{eq:negative-moment-mse-expansion}.
\end{proof}

The above negative-moment condition can also be understood as a requirement on the denominator proposal $q_2$. Namely, the expectation
$$
\mathbb{E}[V^{-\alpha}]
=
\int_{\mathcal{X}}
\frac{q_2(\boldsymbol{x})^{1+\alpha}}{\pi(\boldsymbol{x})^\alpha}\,d\boldsymbol{x}
$$
 may be infinite when $q_2$ is much heavier-tailed than $\pi$, because samples in those tails produce very small denominator weights.

\section{Generality of the joint construction}\label{app:joint-generality}
In Section~\ref{sec:construction_joint}, we fix the marginal transports and choose how to couple their uniform inputs. Under the conditions below, fixing these maps still allows every joint proposal with the prescribed marginals. The vector Sklar theorem also shows that its coupling of uniform inputs is unique.

\begin{proposition}[Vector Sklar theorem, {\citealp[Proposition~1 and Theorem~1]{fan2023vector}}]
Let $\mathbb Q_1,\mathbb Q_2$ be absolutely continuous distributions on $\mathbb R^{\dimens}$ with support contained in convex sets. For $i=1,2$, let $\boldsymbol T_i=\nabla\psi_i$ transport $\mathrm{Unif}([0,1]^{\dimens})$ to $\mathbb Q_i$, where $\psi_i$ is convex. Let $\psi_i^*$ denote its convex conjugate and $\boldsymbol R_i=\nabla\psi_i^*$ the almost-everywhere inverse of $\boldsymbol T_i$. Every joint distribution $\mathbb Q_{1:2}$ with these marginals has a unique coupling $\mathbb C_{1:2}$ with uniform marginals such that
$$
\mathbb Q_{1:2}=(\boldsymbol T_1,\boldsymbol T_2)\sharp\mathbb C_{1:2}.
$$
It is given by $\mathbb C_{1:2}=(\boldsymbol R_1,\boldsymbol R_2)\sharp\mathbb Q_{1:2}$.
\end{proposition}
\begin{proof}
Proposition~1 of \citet{fan2023vector} gives the inverse maps and their pushforward properties under these assumptions. Given a pair with distribution $\mathbb Q_{1:2}$, we apply $\boldsymbol R_i$ to its $i$th component. Each resulting component is uniform, so their joint distribution is an admissible $\mathbb C_{1:2}$. Applying $\boldsymbol T_1$ and $\boldsymbol T_2$ takes us back to the original pair almost surely. The representation is unique because applying the inverse maps to any other proposed representation would give the same distribution $(\boldsymbol R_1,\boldsymbol R_2)\sharp\mathbb Q_{1:2}$.
\end{proof}
The result is very general. For instance it does not require the joint to have a density, again. It was derived in a context of statistical modelling, where making assumptions about marginals can be debatable. In our context, the marginals are IS proposals that we get to choose.

\section{Details of the US Crime experiment}\label{app:uscrime-coupling}
In the US Crime experiment of Section~\ref{sec:experiments}, we used Gaussian regression with $n=47$ observations and $d=58$ coefficients:
$\boldsymbol y\mid\boldsymbol\beta,v\sim\mathcal N_n(\boldsymbol X\boldsymbol\beta,v\boldsymbol I_n)$, $\boldsymbol\beta\sim\mathcal N_d(\boldsymbol0,\boldsymbol\Sigma)$, and $v\sim\operatorname{BP}(5/2,5/2)$. We study how the posterior variance of $G=\boldsymbol\beta^{\mathsf T}\boldsymbol\Sigma^{-1}\boldsymbol\beta/(dv)$ changes when the prior median of $v$ is multiplied by $e^\alpha$. Let $\pi_\alpha(\boldsymbol\beta,v)$ denote the joint posterior, with $\pi=\pi_0$, and let $\pi_v$ be the marginal posterior density of $v$ at $\alpha=0$. The derivative of the log prior density at zero is $s(v)=5/2-5/(1+v)$. With $m=\mathbb{E}_{\pi(\boldsymbol\beta,v)}G$ and $a=\mathbb{E}_{\pi_v}s(v)$, the quantity of interest is 
$$
\theta=\left.\partial_\alpha\operatorname{Var}_{\pi_\alpha(\boldsymbol\beta,v)}(G)\right|_{\alpha=0}
=\mathbb{E}_{\pi(\boldsymbol\beta,v)}[\varphi],\qquad
\varphi(\boldsymbol\beta,v)=(G-m)^2\{s(v)-a\}.
$$
Small values of $v$ can make $G$, and hence $|\varphi(\boldsymbol\beta,v)|$, large. The score $s(v)$ itself is bounded. At the end of this section we verify that $\varphi\in L^1(\pi)\setminus L^2(\pi)$, so direct iid MC has infinite variance.

\paragraph{Choosing the proposals.}
Conditional on $v$, the posterior of $\boldsymbol\beta$ is Gaussian, $\mathcal N_d(\boldsymbol m_v,\boldsymbol C_v)$. We keep this conditional distribution in $q_1$ and change how often we sample each value of $v$. Define
$\rho(v)=\{\mathbb{E}_{\pi(\boldsymbol\beta\mid v,\boldsymbol y)}[\varphi(\boldsymbol\beta,v)^2]\}^{1/2}$ and $K=\int_0^\infty\pi_v(v)\rho(v)\,dv$. For a density $g$ on $(0,\infty)$, consider the proposal $g(v)\pi(\boldsymbol\beta\mid v,\boldsymbol y)$. Writing $\widetilde\pi$ for the unnormalized posterior and $Z_\pi$ for its normalizing constant, Cauchy--Schwarz bounds the second moment of the numerator importance weight by
$$
Z_\pi^2\int_0^\infty\frac{\pi_v(v)^2\rho(v)^2}{g(v)}\,dv
\geq Z_\pi^2\left\{\int_0^\infty\pi_v(v)\rho(v)\,dv\right\}^{2}
=Z_\pi^2K^2.
$$
Equality holds for $g(v)=\pi_v(v)\rho(v)/K$, which therefore minimizes this second moment among proposals with the chosen conditional distribution. We verify $K<\infty$ below. Adding a small posterior component gives our numerator proposal, while for the denominator we slightly broaden the conditional Gaussian:
$$
\begin{aligned}
q_1(\boldsymbol\beta,v)
&=\pi(\boldsymbol\beta,v)\{0.05+0.95\rho(v)/K\},\\
q_2(\boldsymbol\beta,v)
&=0.95\,\pi_v(v)\mathcal N_d(\boldsymbol\beta;\boldsymbol m_v,1.10\boldsymbol C_v)
  +0.05q_1(\boldsymbol\beta,v).
\end{aligned}
$$
The posterior component in $q_1$ covers zeros of $\rho$. The $q_1$ component in $q_2$ gives more probability to the small-$v$ region than the posterior does. Since $q_1\geq0.95\pi\rho/K$ and $q_2\geq0.05q_1$, both proposals give finite second moments for the numerator importance weights.

\paragraph{Constructing the coupling.}
We first construct fixed maps $T_1,T_2$ that transform a standard Gaussian input into samples from $q_1,q_2$. We then learn a transformation $\boldsymbol C_{\boldsymbol\psi}$ that preserves the standard Gaussian distribution and use
$$
(\boldsymbol\beta_1,v_1)=T_1(\boldsymbol Z),\qquad
(\boldsymbol\beta_2,v_2)=T_2(\boldsymbol C_{\boldsymbol\psi}(\boldsymbol Z)),
\qquad \boldsymbol Z\sim\mathcal N_{59}(\boldsymbol0,\boldsymbol I_{59}).
$$
Thus changing $\boldsymbol\psi$ changes the dependence while keeping $q_1,q_2$ fixed. All 59 variables are sampled. The learned transformation acts on four coordinates chosen to capture variation in $v$ and $G$, as we explain below.

To define $T_1,T_2$, it helps to express the Gaussian conditional in independent coordinates. Take the singular value decomposition
$$
\boldsymbol X\boldsymbol\Sigma^{1/2}
=\boldsymbol U[\operatorname{diag}(\kappa_1,\ldots,\kappa_n)\ \boldsymbol0]\boldsymbol V^{\mathsf T},
\qquad
\boldsymbol\eta=\boldsymbol V^{\mathsf T}\boldsymbol\Sigma^{-1/2}\boldsymbol\beta,
$$
where $\boldsymbol\Sigma^{1/2}$ is the symmetric square root and $\boldsymbol U,\boldsymbol V$ are orthogonal. Full row rank of $\boldsymbol X$ gives $\kappa_j>0$. With $\boldsymbol b=\boldsymbol U^{\mathsf T}\boldsymbol y$, the conditional means and variances of the independent coordinates $\eta_j$ are
$$
(\boldsymbol\mu_v)_j=\frac{\kappa_jb_j}{\kappa_j^2+v},\quad
(\boldsymbol c_v)_j=\frac{v}{\kappa_j^2+v}\quad(j\leq n),
\qquad
(\boldsymbol\mu_v)_j=0,\quad(\boldsymbol c_v)_j=1\quad(j>n).
$$
The likelihood involves only the first $n$ coordinates. The last $d-n=11$ retain their prior distribution, independently of $v$. In these coordinates $G=\|\boldsymbol\eta\|^2/(dv)$, and we recover $\boldsymbol\beta=\boldsymbol\Sigma^{1/2}\boldsymbol V\boldsymbol\eta$.

Write $\boldsymbol Z=(\boldsymbol g,z_v)$ and let $\Phi$ denote the standard normal CDF. Let $q_{j,v}$ and $F_{q_{j,v}}$ be the density and CDF of $v$ under $q_j$. Then $T_1$ uses $v=F_{q_{1,v}}^{-1}(\Phi(z_v))$ and $\boldsymbol\eta=\boldsymbol\mu_v+\sqrt{\boldsymbol c_v}\odot\boldsymbol g$, where the square root and product act coordinatewise.

For $T_2$, we also want a continuous transformation so that we can use pathwise gradients. In $\boldsymbol\eta$ coordinates, subtracting $\boldsymbol\mu_v$ and dividing coordinatewise by $\sqrt{\boldsymbol c_v}$ gives two centered Gaussian components with the same uniform direction. We can therefore sample their mixture by changing only the length of $\boldsymbol g$. The conditional CDF of the squared length is
$$
F_v(t)=w(v)F_{\chi^2_d}(t/1.10)+\{1-w(v)\}F_{\chi^2_d}(t),
\qquad
w(v)=\frac{0.95\pi_v(v)}{q_{2,v}(v)},
$$
where $q_{2,v}=0.95\pi_v+0.05q_{1,v}$. We set $v=F_{q_{2,v}}^{-1}(\Phi(z_v))$ and
$$
\boldsymbol\eta=\boldsymbol\mu_v+\sqrt{\boldsymbol c_v}\odot
\frac{\boldsymbol g}{\|\boldsymbol g\|}
\sqrt{F_v^{-1}\!\left(F_{\chi^2_d}(\|\boldsymbol g\|^2)\right)}.
$$
Transforming back to $\boldsymbol\beta$ gives a sample from the full mixture $q_2$.

\paragraph{The four coordinates.}
To choose where the learned transformation should act, consider the expression for $G$ under $T_1$:
$$
\|\boldsymbol\eta\|^2
=\|\boldsymbol\mu_v\|^2
+2(\boldsymbol\mu_v\odot\sqrt{\boldsymbol c_v})^{\mathsf T}\boldsymbol g
+\sum_{j=1}^d(\boldsymbol c_v)_jg_j^2.
$$
Alongside $z_v$, we use the length of $\boldsymbol g$ and its angle with a direction aligned with the linear term. We fix this direction at the posterior median $v_\star$ of $v$:
$\boldsymbol e=(\boldsymbol\mu_{v_\star}\odot\sqrt{\boldsymbol c_{v_\star}})/\|\boldsymbol\mu_{v_\star}\odot\sqrt{\boldsymbol c_{v_\star}}\|$, whose denominator is nonzero for these data.
The final coordinate describes how much of $\boldsymbol g$ lies in the last 11 coordinates, whose conditional variances stay at one when $v$ is small.

Write $\boldsymbol g=(\boldsymbol g_o,\boldsymbol g_n)\in\mathbb R^n\times\mathbb R^{d-n}$ and $\boldsymbol e=(\boldsymbol e_o,\boldsymbol0)$. Define
$$
R=\|\boldsymbol g\|,\qquad
c=\frac{\boldsymbol e^{\mathsf T}\boldsymbol g}{R},\qquad
h=\frac{\|\boldsymbol g_n\|^2}
 {\|\boldsymbol g_o-Rc\boldsymbol e_o\|^2+\|\boldsymbol g_n\|^2}.
$$
Here $h$ is the fraction of squared length in $\boldsymbol g_n$ after removing the component along $\boldsymbol e$. Gaussian symmetry gives independent $R,c,h$, with $R^2\sim\chi^2_d$, $h\sim\operatorname{Beta}((d-n)/2,(n-1)/2)$, and $c$ distributed as one coordinate of a uniform direction on the $(d-1)$-sphere. Writing $F_c,F_h$ for the CDFs of $c,h$, we obtain four independent standard Gaussian coordinates:
$$
\boldsymbol x=\left(z_v,\Phi^{-1}(F_{\chi^2_d}(R^2)),
\Phi^{-1}(F_c(c)),\Phi^{-1}(F_h(h))\right)
\sim\mathcal N_4(\boldsymbol0,\boldsymbol I_4).
$$
We apply an orthogonal $4\times4$ map to $\boldsymbol x$, followed by the conditional rotations from Section~\ref{app:uncon} on coordinates $(3,4),(1,4),(2,3),(1,3),(2,4),(1,2)$, in that order. Each rotation preserves the standard Gaussian distribution, so the resulting $\boldsymbol x'$ has the same distribution as $\boldsymbol x$.

To reconstruct the full input, we retain the directions
$\boldsymbol u_o=(\boldsymbol g_o-Rc\boldsymbol e_o)/\|\boldsymbol g_o-Rc\boldsymbol e_o\|$ and $\boldsymbol u_n=\boldsymbol g_n/\|\boldsymbol g_n\|$, which are uniform on their respective unit spheres and independent of $\boldsymbol x$. Inverting the CDF transformations gives $R',c',h'$, and we set
$$
\boldsymbol g'_o=R'\{c'\boldsymbol e_o+
\sqrt{(1-c'^2)(1-h')}\boldsymbol u_o\},\qquad
\boldsymbol g'_n=R'\sqrt{(1-c'^2)h'}\boldsymbol u_n,\qquad z'_v=x'_1.
$$
The retained directions are also independent of $\boldsymbol x'$. Thus $\boldsymbol C_{\boldsymbol\psi}(\boldsymbol Z)=(\boldsymbol g',z'_v)$ is standard Gaussian in all 59 dimensions, as required to preserve $q_2$.

\paragraph{Training and selecting the coupling.}
An independent pilot gave a negative estimate of $\theta$. Following the objective in Section~\ref{sec:optimalcoupling}, we used Adam and pathwise gradients to minimize
$$
\mathbb{E}_{\boldsymbol Z}\!\left[
\frac{\varphi(\boldsymbol\beta_1,v_1)\widetilde\pi(\boldsymbol\beta_1,v_1)}{q_1(\boldsymbol\beta_1,v_1)}
\frac{\widetilde\pi(\boldsymbol\beta_2,v_2)}{q_2(\boldsymbol\beta_2,v_2)}
\right].
$$
We used six orthogonal matrices, learned during the pilot run. For each matrix, we trained the networks that determine the rotation angles (as explained earlier in the Appendix) from three initial parameter settings, keeping the matrix fixed. We then selected the saved parameters with the smallest estimated cross moment on a separate set of samples. 

\paragraph{Why iid MC has infinite variance.}
Put $\boldsymbol A=\boldsymbol X\boldsymbol\Sigma\boldsymbol X^{\mathsf T}$. Full row rank makes $\boldsymbol A$ positive definite. Integrating out $\boldsymbol\beta$ gives
$$
\pi_v(v)\propto
v^{3/2}(1+v)^{-5}|\boldsymbol A+v\boldsymbol I_n|^{-1/2}
\exp\{-\boldsymbol y^{\mathsf T}(\boldsymbol A+v\boldsymbol I_n)^{-1}\boldsymbol y/2\}.
$$
The determinant and exponential factors approach positive finite values as $v\downarrow0$, so $\pi_v(v)\sim C v^{3/2}$ for some $C>0$. Meanwhile, the last 11 coordinates of $\boldsymbol\eta$ retain unit variance. As $v\downarrow0$, $Q=\|\boldsymbol\eta\|^2$ approaches in distribution
$\sum_{j=1}^n(b_j/\kappa_j)^2+\chi^2_{d-n}$, and its conditional second and fourth moments approach positive finite values. Thus $G=Q/(dv)$ grows on the order of $v^{-1}$. Its posterior mean $m$ is nevertheless finite: near zero, the integrand for $\mathbb{E}_{\pi(\boldsymbol\beta,v)}G$ has order $v^{1/2}$. Also $s(v)\to-5/2$, whereas $a$ lies strictly between $-5/2$ and $5/2$. From $\varphi=(G-m)^2\{s(v)-a\}$ we obtain
$$
\mathbb{E}_{\pi(\boldsymbol\beta\mid v,\boldsymbol y)}[|\varphi|]\asymp v^{-2},\qquad
\mathbb{E}_{\pi(\boldsymbol\beta\mid v,\boldsymbol y)}[\varphi^2]\asymp v^{-4}\qquad(v\downarrow0).
$$
After multiplication by $\pi_v(v)$, these have orders $v^{-1/2}$ and $v^{-5/2}$. Only the first is integrable at zero. For $v\geq1$, the score and the conditional moments of $Q$ are bounded, so $\rho(v)$ is bounded too. Hence $\mathbb{E}_{\pi(\boldsymbol\beta,v)}|\varphi|<\infty$ while $\mathbb{E}_{\pi(\boldsymbol\beta,v)}\varphi^2=\infty$.

We can now also check that the numerator proposal is well defined: $\rho(v)\asymp v^{-2}$ near zero, so $\pi_v(v)\rho(v)\asymp v^{-1/2}$ is integrable and $K<\infty$.

\section*{Expressive marginals: normalizing flows with input-convex NNs}\label{app:normalizing-flows}
 A crucial issue in IS is the specification of parametric proposal distributions that can work in high-dimensional settings. For this purpose, normalizing flows (NFs) have shown promise \citep{muller2019neural}. NFs define a diffeomorphic transport map between an arbitrary initial distribution (such as uniform) and a proposal, in our case. In fact, \citep{DBLP:conf/iclr/HuangCTC21} develop NF architectures that are Brenier maps, i.e., they directly fit in the framework we described above.

\medskip
\bibliography{references}
\end{document}